\documentclass[aps,prb,reprint,superscriptaddress]{revtex4-1}

\usepackage{graphicx}
\usepackage{bm}
\usepackage{multirow}
\usepackage{amsmath}

\begin{document}

\title{Anomalous angular dependence of the upper critical induction of orthorhombic ferromagnetic superconductors with completely broken \emph{p}-wave symmetry}


\author{Christopher L\"{o}rscher }

\affiliation{Department of Physics, University of Central Florida, Orlando, FL 32816-2385 USA}
\author{Jingchuan Zhang}
\affiliation{Department of Physics, University of Central Florida, Orlando, FL 32816-2385 USA}
\affiliation{Department of Physics, University of Science and Technology Beijing, Beijing 100083, China}
\author{Qiang Gu}
\affiliation{Department of Physics, University of Science and Technology Beijing, Beijing 100083, China}
\author{Richard A. Klemm}
\affiliation{Department of Physics, University of Central Florida, Orlando, FL 32816-2385 USA}


\date{\today}

\begin{abstract}
We employ the Klemm-Clem transformations to map the equations of motion for the Green functions of a clean superconductor with a general ellipsoidal Fermi surface (FS) characterized by the effective masses $m_1, m_2$, and $m_3$ in the presence of an arbitrarily directed magnetic induction ${\bm B}=B(\sin\theta\cos\phi,\sin\theta\sin\phi,\cos\theta)$ onto those of a spherical FS. We then obtain the transformed gap equation for a transformed pairing interaction $\tilde{V}(\hat{\tilde{\bm k}},\hat{\tilde{\bm k}}')$ appropriate for any orbital order parameter symmetry.
We use these results to calculate the upper critical induction $B_{c2}(\theta,\phi)$ for an orthorhombic ferromagnetic superconductor with transition temperatures $T_{\rm Curie}>T_c$. We assume the FS is split by strong spin-orbit coupling, with a single parallel-spin ($\uparrow\uparrow$) pairing interaction of the \textit{p}-wave polar state form locked onto the $\hat{\bm e}_3$ crystal axis normal to the spontaneous magnetization ${\bm M}_0\perp\hat{\bm e}_3$ due to the ferromagnetism. The orbital harmonic oscillator eigenvalues are modified according to $B\rightarrow B\alpha$, where $\alpha(\theta,\phi)=\sqrt{m_3/m}\sqrt{\cos^2\theta+\gamma^{-2}(\phi)\sin^2\theta}$, $\gamma^2(\phi)=m_3/(m_1\cos^2\phi+m_2\sin^2\phi)$ and $m=(m_1m_2m_3)^{1/3}$. At fixed $\phi$, the order parameter anisotropy causes $B_{c2}$ to exhibit a novel $\theta$-dependence, which for $\gamma^2(\phi)>3$ becomes a double peak at $0^{\circ}<\theta^{*}<90^{\circ}$ and at $180^{\circ}-\theta^{*}$, providing a sensitive bulk test of the order parameter orbital symmetry in both phases of URhGe and in similar compounds still to be discovered.

\end{abstract}


\maketitle

\section{Introduction}
Recent discoveries of materials with coexistent superconductivity and ferromagnetism and of superconducting doped topological insulators have renewed interest in parallel-spin triplet superconductivity, the simplest cases having \textit{p}-wave orbital symmetry\cite{Huy,deVisser,Aoki,HH,Levy,Levy2,Yelland,Aoki2,AokiFlouquet,SK1980,SK1985,KS,Mineev,Davis,Shick,Mueller,VG, Blount,Sauls,Shivaram,ChoiSauls,MachidaMachida,MS,MM,Maeno,Deguchi,Kittaka,Yonezawa,Suderow,Machida,Kriener,BayTI}. Ferromagnetic superconductors have the ferromagnetic transition temperature $T_{\rm Curie}$ exceeding the superconducting transition temperature $T_c$. In ferromagnetic superconductors, one can measure the temperature $T$ and orientation dependence of the upper critical field ${\bm H}_{c2}$, at which the superconductivity is destroyed by the applied magnetic field ${\bm H}$ in combination with the ferromagnetic spontaneous magnetization ${\bm M}_0$. However, in such materials, it is more convenient to calculate the upper critical magnetic induction ${\bm B}_{c2}$, which arises from the complicated interplay of ferromagnetic and diamagnetic superconducting components in the single function ${\bm B}=\mu_0{\bm H}+{\bm M}$, where ${\bm M}({\bm H})$ is the field-dependent magnetization. One can probe the bulk properties of the superconductivity by measuring the
 $T$ and differently oriented ${\bm H}$ dependencies of ${\bm B}_{c2}$\cite{SK1980,SK1985,KS,VG,Blount,Sauls,MS}. Ambient pressure measurements of the bulk probe ${\bm B}_{c2}$ and of local probes such as muon depolarization experiments of orthorhombic UCoGe \cite{Huy,deVisser} and URhGe\cite{Aoki,HH,Levy,Levy2,Yelland} showed
that the superconductivity exists completely within the ferromagnetic $T$ range and that the same electrons are responsible for the superconductivity and the ferromagnetism \cite{deVisser,AokiFlouquet}. In some non-ferromagnetic $p$-wave superconductors, such as the purported doped topological insulators, although ${\bm M}_0=0$, there can still be complications due to competing surface and bulk properties. The variety of possible \textit{p}-wave
states can still be characterized in those materials by bulk measurements of ${\bm H}_{c2}(T)$ for a variety of ${\bm H}$ orientations.

The orbital symmetry of the superconducting order parameter usually can be classified by its nodes both in the order parameter and in the resulting superconducting energy gap. For $p$-wave superconductors free of long-range ferromagnetism, one may have a nodeless gap, such as for the isotropic Balian-Werthamer (BW) state of $^3$He \cite{BW}, or a gap with either planar nodes (polar state), or point nodes (axial state), where it vanishes on the Fermi surface (FS). The basic order parameter symmetries of these three basic order parameters are depicted in Fig. 1. Each of these states possesses unique $T$ and $\hat{\bm H}$ orientational dependencies of ${\bm H}_{c2}(T)$, which are useful in identifying the orbital symmetries experimentally. It was shown theoretically by Scharnberg and Klemm that for $p$-wave superconductors with an isotropic equal-spin pairing interaction of the form $V_{3D}(\hat{\bm k},\hat{\bm k}')=V_{0}\hat{\bm k}\cdot\hat{\bm k}'$, which leads to an isotropic BW state for ${\bm H}=0$ with an isotropic gap function as sketched in Fig. 1(a), ${\bm H}_{c2}(T)$ is always given by that of the polar state, $H_{c2,{\rm polar}}(T)$\cite{SK1980}, in which ${\bm H}$ always points in an antinodal order parameter direction. This is analogous to the interaction of ${\bm H}$ with spins through the rotationally-invariant Heisenberg interaction with an isotropic ${\bm g}$-tensor. To avoid confusion with the various order parameter states, we hereby designate $H_{c2,{\rm p}\>{\rm antinodal}}(T)\equiv H_{c2,{\rm polar}}(T)$. Except for the $p$-wave chiral ABM states\cite{Zhang}, when ${\bm H}$ lies along the antinodal direction, $H_{c2}(T)=H_{c2,{\rm p}\>{\rm antinodal}}(T)$, even though the state symmetry may be very different than that of the polar state. $H_{c2,{\rm p}\>{\rm antinodal}}(T)$ has a much straighter $T$ dependence than any other $p$-wave or $s$-wave state in pure, three-dimensional materials with a spherical (or ellipsoidal, as shown here) FS\cite{SK1980}. Although one might question the notion of an isotropic $p$-wave pairing interaction in a layered superconductor\cite{book}, the apparent presence of a rather isotropic gap in the doped topological insulator, Cu$_x$Bi$_2$Se$_3$\cite{Kriener}, led the de Visser group to investigate ${\bm H}_{c2}(T)$ both $||$ and $\perp$ to the Bi$_2$Se$_3$ layers, and they found good agreement with the appropriately scaled $H_{c2,{\rm p}\>{\rm antinodal}}(T)$ in both directions\cite{BayTI,SK1980}.

\begin{figure}
\center{\includegraphics[width=0.9\linewidth]{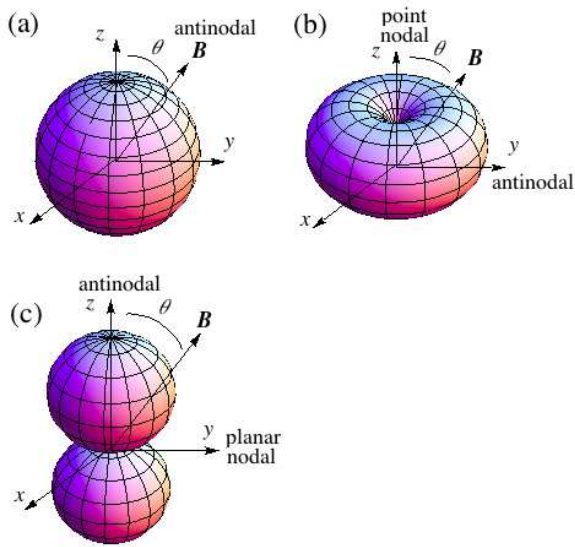}
\caption{Sketches of the three basic types of $p$-wave gap functions $|\Delta(\hat{\bm k})|$. (a) The non-chiral BW, or isotropic gap $|\Delta_0|$ $p$-wave state, for which $H_{c2}(T)$ is given by $H_{c2,{\rm p}\>{\rm antinodal}}(T)$ for all ${\bm H}$ directions\cite{SK1980}. (b) The ABM and SK states. When these states have their antinodal planes locked onto a uniaxial crystal plane, breaking the planar antinodal axial rotational symmetry, the chiral ABM states have complex order parameters $\Delta_{0\pm}(\hat{k}_x\pm i\hat{k}_y)$ with distinct $H_{c2,{\rm ABM}\>{\rm nodal}}(T)$ and $H_{c2,{\rm ABM}\>{\rm antinodal}}(T)$ for ${\bm H}$ along the nodal axis and antinodal planar directions, respectively\cite{SK1980,Zhang}. The SK state with order parameter $\sum_{\sigma=\pm}\Delta_{0,\sigma}(\hat{k}_x+i\sigma\hat{k}_y)$ is more complicated. For ${\bm H}$ along the nodal axis, the SK state is chiral with $H_{c2,{\rm SK}\>{\rm nodal}}(T)$\cite{SK1980}. For ${\bm H}$ in the antinodal plane, the SK state is non-chiral with $H_{c2,{\rm p}\>{\rm antinodal}}(T)$\cite{Zhang}. See text. (c) The non-chiral polar/CBS state. This state with order parameter $\Delta_0 k_z$ has its antinodal axis locked onto a crystal axis (e.g., the $\hat{\bm z}$ axis), breaking the point antinodal axial rotational symmetry. For ${\bm H}$ parallel and perpendicular to the antinodal axis, $H_{c2}(T)$ is respectively $H_{c2,{\rm p}\>{\rm antinodal}}(T)$ and the distinct planar nodal form, $H_{c2,{\rm planar}\>{\rm nodal}}(T)$ \cite{SK1985}.}}
\end{figure}\label{fig1}

Scharnberg and Klemm also investigated the effects of two pairing states perpendicular to ${\bm H}$ within the framework of the rotationally symmetric $V_{3D}(\hat{\bm k},\hat{\bm k}')$. For ${\bm H}||\hat{\bm z}$, there are two order parameter components, which are usually written as $\Delta_{\pm}(\hat{\bm k})=\Delta_{\pm,0}( \hat{k}_x\pm i\hat{k}_y)$, both components of which nominally share the same $T_c$.
 These are the two chiral manifestations of the Anderson-Brinkman-Morel (ABM) state of $^3$He\cite{AM,AB}, in which only parallel-spin pairing with one spin state is involved. These ABM states with ${\bm H}=0$ have a gap function with a nodal point, as sketched in Fig. 1(b). Scharnberg and Klemm also investigated $H_{c2}(T)$ for the special case of ${\bm H}$ along the nodal point direction normal to the pairing plane of these chiral ABM states, and found that $H_{c2,{\rm ABM}\>{\rm nodal}}(T)$ for either of these ABM states exhibited a $T$ dependence that rose even more slowly with decreasing $T$ than did $H_{c2,s}(T)$ for a pure, isotropic $s$-wave superconductor on a spherical (or ellipsoidal, as shown here) FS in the absence of Pauli-limiting effects\cite{SK1980}.

 However, Scharnberg and Klemm then investigated the effects of the two combined chiral ABM pairing states perpendicular to ${\bm H}$. In effect, they calculated $H_{c2}(T)$ for the two-component state containing an unequal amplitude mix of the two chiral ABM states, $\Delta_{\rm SK}(\hat{\bm k})=\sum_{\sigma=\pm}\Delta_{0,\sigma}(\hat{k}_x+i\sigma\hat{k}_y)$\cite{SK1980}. The SK state is a chiral state except for the special cases when $|\Delta_{0,+}|=|\Delta_{0,-}|=\Delta_0$, for which it is non-chiral. For those special cases, one may write $\Delta_{\rm SK}(\hat{\bm k})=\Delta_0\sum_{\sigma=\pm}e^{i\psi_{\sigma}}(\hat{k}_x+i\sigma\hat{k}_y)$, which may be rewritten as $\Delta_{\rm SK}(\hat{\bm k})=2\Delta_0e^{i\phi_{+}}\sin\theta_{\bm k}\cos(\phi_{\bm k}+\phi_{-})$, where $\phi_{\pm}=(\psi_{+}\pm\psi_{-})/2$ is independent of $\hat{\bm k}$. Except for the overall constant phase $\phi_{+}$, $\Delta_{\rm SK}(\hat{\bm k})$ is therefore a real function of $\hat{\bm k}$ and hence non-chiral whenever $|\Delta_{0,+}|=|\Delta_{0,-}|$. The magnetic analog of this degenerate, two-component state is the anisotropic XY model of spin-spin interactions, in which there is an easy plane normal to a hard axis for spin-spin interactions with ${\bm H}$ in that plane, but the interactions within the easy plane can be either isotropic or anisotropic, depending upon the field direction. Although they originally denoted this as the ``generalized ABM state''\cite{SK1980}, this state came to be known as the SK state\cite{SK1985,LM,MS}. For ${\bm H}||\hat{\bm z}$, the chiral SK state has $H_{c2,{\rm SK}\>{\rm nodal}}(T)$. However, for ${\bm H}\perp\hat{\bm z}$, the SK state is non-chiral just below $H_{c2,{\rm p}\>{\rm antinodal}}(T)$\cite{Zhang}. The precise form of $H_{c2,{\rm planar}\>{\rm nodal}}(T)$ and the interesting transition from chiral to non-chiral signatures in $H_{c2}(\theta,T)$ for the SK state at precise intermediate $\theta$ values will be presented elsewhere\cite{Zhang}. Although not mentioned in the original paper\cite{SK1980}, the SK and ABM states might be favored in superconductors with uniaxial symmetry such as certain layered superconductors\cite{book}, for which $V_{2D}(\hat{\bm k},\hat{\bm k}')=V_0(\hat{k}_x\hat{k}_x'+\hat{k}_y\hat{k}_y')$ could lock onto the layers, breaking the axial rotational degree of freedom of the antinodal plane. Sr$_2$RuO$_4$ has often been mentioned as a likely candidate for either the single parallel-spin chiral ABM state or the dual parallel-spin SK state, which is either chiral or non-chiral, depending upon the direction of ${\bm H}$, although many of the authors were apparently unaware of the proper designation of the latter state they described\cite{MM,Maeno}. For the ABM state, when ${\bm H}$ is parallel to the antinodal plane, $H_{c2}(T)$ is given by the new form $H_{c2,{\rm ABM}\>{\rm antinodal}}(T)$\cite{Zhang}. Neither the ABM nor the SK state appears to be consistent with the experiments of $H_{c2,||}(T)$ parallel to the layers of Sr$_2$RuO$_4$\cite{Deguchi,Kittaka,Yonezawa}, which show that $H_{c2,||}(T)$ is strongly Pauli limited\cite{Machida,book}. Recent scanning tunneling microscopy on that material were also inconsistent with gap nodes\cite{Suderow}. Regardless of whether Sr$_2$RuO$_4$ or some other as yet undiscovered material will be the first manifestation of the SK or ABM states, ${\bm H}_{c2}(\theta)(T)$ at an arbitrary angle $\theta$ with respect to the fixed nodal point direction of the SK or ABM states with the normal state electrons on a general ellipsoidal FS will be presented elsewhere\cite{Zhang}.

 Finally, the case of particular interest in this paper is that of an anisotropic $p$-wave pairing interaction with equal-spin pairing along only one direction, the one-dimensional (1D) analog of $V_{3D}(\hat{\bm k},\hat{\bm k}')$, or $V_{1D}(\hat{\bm k},\hat{\bm k}')=V_0\hat{k}_z\hat{k}_z'$ \cite{SK1985}. This state, $\Delta_0 k_z$, has come to be known as the polar/CBS state, for a polar state of completely broken rotational symmetry, analogous to the Ising interaction representing the dominant easy-axis component of the highly anisotropic 3D Heisenberg spin-spin interaction. A sketch of the polar/CBS gap function is given in Fig. 1(c). As for the ABM or SK superconducting states in a crystal, the 1D pairing is fixed to the crystal lattice, but in this case, to one crystal axis direction only. The largest intrinsic anisotropy due solely to the order parameter arises between the field applied parallel and perpendicular to this single pairing direction. If the field is along the pairing or antinodal direction, as in the 3D case, one obtains $H_{c2,{\rm p}\>{\rm antinodal}}(T)$\cite{SK1980}. However, when the field is applied in the planar nodal direction perpendicular to the pairing, then $H_{c2}(T)$ has a distinctly different form, $H_{c2,{\rm planar}\>{\rm nodal}}(T)$, similar to but not identical to $H_{c2,s}(T)$ \cite{SK1985}. Summarizing the various cases evaluated prior to this work, we have for all $T$ with pairing on spherical FSs\cite{SK1980,SK1985,Zhang},
\begin{align*}
	&H_{c2,\text{p antinodal}} > H_{c2,\text{SK nodal}} > H_{c2,\text{ABM antinodal}}\notag\\
	&\qquad\qquad> H_{c2,s} > H_{c2,\text{planar nodal}} > H_{c2,\text{ABM nodal}}.
\end{align*}

 The angular dependence of either ${\bm H}_{c2}(T)$ or ${\bm B}_{c2}(T)$ for the 1D polar/CBS state case is important to aid experimentalists in determining its realization in materials such as URhGe. These new results are the focus of this paper. Since URhGe, the existing material for which this polar/CBS state has been strongly supported by experiment \cite{HH}, has an orthorhombic crystal structure \cite{Davis,Shick,Mueller,AokiFlouquet}, its FS can be approximated as a general ellipsoid. Although the critical field data of UCoGe are more suggestive of an SK or ABM state at low ${\bm H}$ values, its crystal structure is also orthorhombic \cite{AokiFlouquet}. Hence, we have derived the prescription for including general ellipsoidal FS anisotropies into microscopic calculations of $B_{c2}(T)$ for a general anisotropic pairing interaction $V(\hat{\bm k},\hat{\bm k}')$, and with the magnetic induction ${\bm B}$ in a general direction. The details of the derivation are presented in the appendix. In this paper, we used this procedure to calculate the full angular dependence of $B_{c2}(\theta,\phi,T)$ for the polar/CBS state of a ferromagnetic superconductor dominated by a single parallel-spin state, and our results are presented.

In the extraordinary case of URhGe,
${\bm B}_{c2}(T)$ measurements on a sample with a residual resistance ratio (RRR) = 21
were fit to the Scharnberg-Klemm theory of the $p$-wave polar/CBS state along all three crystallographic directions, with equal spin pairing along the $a$-axis direction and weak ferromagnetism along the $c$-axis direction in the low-field regime, using the resistively measured slopes of $B_{c2}$ along the $a$-, $b$-, and $c$-axis directions just below the ferromagnetic demagnetization jumps at $T_c$ as the only
fitting parameters\cite{HH}. The measured $B_{c2,a}(T)$ fit the predicted $H_{c2,{\rm p}\>{\rm antinodal}}(T)$ behavior, but $B_{c2,b}(T)$ and $B_{c2,c}(T)$ fit the qualitatively different $H_{c2,{\rm planar}\>{\rm nodal}}(T)$ curve\cite{SK1985}, with a constant ratio $B_{c2,b}(T)/B_{c2,c}(T)$ consistent with $T$-independent FS anisotropy. $B_{c2}(0)$ in all three crystal directions violated the Pauli limit $B_P\sim 1.85 T_c$ T/K for a singlet-spin $s$-wave superconductor \cite{HH}, indicating that URhGe is very unlikely to be an $s$- or $d$-wave superconductor. Consequently, these data provided strong evidence that
the superconducting order parameter is likely to
 have the simplest parallel-spin $p$-wave orbital form $\hat{\bm d}k_a$ consistent with ferromagnetism in the $bc$ plane of an orthorhombic crystal, where the pair-spin vector $\hat{\bm d}=(\hat{\bm b}+i\hat{\bm c})/\sqrt{2}$, and the $p$-wave pairing interaction fixed to the crystal $a$-axis direction for all ${\bm M}({\bm H})\perp\hat{\bm a}$ directions and the two possible parallel-spin states indicated by $\hat{\bm b}=|\uparrow\uparrow\rangle$ and $\hat{\bm c}=|\downarrow\downarrow\rangle$ \cite{Mineev}.

Subsequent measurements on a URhGe
 sample with RRR = 50 \cite{Levy} observed an anomalous high ${\bm H}||\hat{\bm b}$ reentrant superconducting phase\cite{Levy}, further supporting the idea of a $p$-wave parallel spin state. But the low-field regime
$B_{c2}(\theta,\phi=0^{\circ})$ within the $ab$ plane was consistent with ordinary FS
anisotropy, at least within the experimental resolution \cite{Levy}. At first sight, these results appear to be in contradiction with the earlier measurements of $B_{c2}$ in URhGe \cite{HH}.

 Note that these results are different than those obtained from hexagonal UPt$_{3}$, which has antiferromagnetic domains with the magnetic ordering along the $a$-axis direction, and for ${\bm H}\perp\hat{\bm c}$, the resulting $H_{c2,\perp {\bm c}}(T)$ is consistent with that of the $p$-wave polar state\cite{SK1985,Shivaram,ChoiSauls}. For ${\bm H}||\hat{\bm c}$, the $H_{c2,||{\bm c}}(T)$ measurements of Shivaram \textit{et. al.} and the calculations of Choi and Sauls fit that of the polar state with Pauli pair breaking for the anti-parallel spin triplet state\cite{Sauls,Shivaram,ChoiSauls}. UPt$_3$ has three superconducting phases, and appears to contain some amount of all three triplet spin states\cite{Sauls,ChoiSauls,MachidaMachida}.
\section{The Model}

In this paper, we calculate $B_{c2}(\theta,\phi,T)$ for a ferromagnetic superconductor with $T_{\rm Curie}>T_c$ and $p$-wave polar/CBS symmetry. Since all three low-field $B_{c2}(T)$ curves for the RRR = 21 crystal of URhGe have different slopes at $T_{c}$, the simplest possible FS to consider is an ellipsoidal one, with $\epsilon({\bm k})=\sum_{i=1}^3k_i^2/(2m_i)$, having three different single particle effective masses $m_{1}$, $m_{2}$, and $m_{3}$, appropriate for orthorhombic symmetry. We calculate $B_{c2}$ within the $ab$-plane for the RRR = 21 and 50 URhGe crystals, and predict that under some conditions, a non-monotonic $B_{c2}(\theta,\phi)$ curve with a double peak at $0^{\circ} < \theta^{*} < 90^{\circ}$ and $180^{\circ}-\theta^{*}$ at fixed $\phi$ could arise, providing
 a definitive bulk test of the orbital symmetry of the order parameter. Our method is applicable to superconductors of any order parameter symmetry.

For our $B_{c2}$ calculations, we assume the strong spin-orbit interaction splits the FS into two FSs, each with only one spin state $\uparrow$ or $\downarrow$, and neglect the $\downarrow$ FS, as if the material were nearly a half metal. We further assume weak
coupling for a clean homogeneous type-II parallel-spin $\uparrow$ $p$-wave superconductor with effective Hamiltonian \cite{SK1980,SK1985},
\begin{align}
\cal{H} &= \sum_{{\bm k}}a_{{\bm k},\uparrow}^{\dag}[\epsilon({\bm k}-e{\bm A})-\mu_{\uparrow}(B)]a^{}_{{\bm k},\uparrow}\nonumber\\
&\qquad+\frac{1}{2}\sum_{{\bm k},{\bm k}'}a_{{\bm k}',\uparrow}^{\dag}a_{{\bm k},\uparrow}^{\dag}V_{\uparrow\uparrow}(\hat{\bm k},\hat{\bm k}')a^{}_{{\bm k},{\uparrow}}a^{}_{{\bm k}',{\uparrow}},\\
V_{\uparrow\uparrow}(\hat{\bm k},\hat{\bm k}')&=3V_{\uparrow\uparrow,0}\hat{k}_{a}^{}\hat{k}'_{a}\hat{\bm d}\cdot\hat{\bm d}^{*}=3V_{\uparrow\uparrow,0}\hat{k}_{a}^{}\hat{k}'_{a},\label{interaction}
\end{align}
where $e$ is the electronic charge, $\hat{\bm d}$ is the vector representing the $|\uparrow\uparrow\rangle$ pair spin states on the $\uparrow$ FS with chemical potential $\mu_{\uparrow}(B)=\mu-g\mu_BB/2$ including the Zeeman interaction, where $\mu_B$ is the Bohr magneton, $g=2$ is assumed to be isotropic, and unit wave vectors are defined on the ellipsoidal $\uparrow$ FS to be
\begin{align}
\hat{k}_{i}\equiv\frac{\sqrt{2m_{i}}}{\alpha(\theta,\phi)}\frac{\partial}{\partial k_{i}}\sqrt{\epsilon({\bm k})}\Bigr|_{\epsilon({\bm k})=\mu_{\uparrow}(B)},\label{kihat}
\end{align}
 where
\begin{align}
\alpha(\theta,\phi)=[\overline{m}_{1}\sin^2\theta\cos^2\phi&+\overline{m}_{2}\sin^2\theta\sin^2\phi\notag\\
&\qquad\qquad+\overline{m}_{3}\cos^2\theta]^{1/2},\label{alpha}
\end{align}
$\overline{m}_{i}=m_{i}/m$, $m=(m_{1}m_{2}m_{3})^{1/3}$,
 and we set $\hbar=k_B=1$. The ellipsoidal $\uparrow$ FS is assumed to be the best approximation to that FS piece most relevant for the superconductivity that can lead to analytic solutions of ${\bm B}_{c2}$ \cite{Davis,Shick,Mueller}.
The orbital symmetry of the equal-spin pairing interaction is that of a \textit{p} wave locked onto the $\hat{\bm a}\equiv\hat{\bm e}_3$ axis of an orthorhombic crystal with ${\bm M}_0||\hat{\bm c}$ on an ellipsoidal FS containing single-particle effective masses $m_{i}$ along the orthogonal $\hat{\bm e}_i$ directions, respectively \cite{Mineev}. The presence of $\alpha(\theta,\phi)$ in Eq.~(\ref{kihat}) is necessary to insure that the transformed unit wave vectors are normal to the transformed spherical $\uparrow$ FS, and that $T_c$ does not depend upon the direction of ${\bm B}$ when ${\bm B}=0$. Here $\alpha(\theta,\phi)$ contains the same effective mass directional dependencies as does the anisotropic Ginzburg-Landau (AGL) model \cite{KC,book}, although the $m_{i}$ in this model differ in principle from the analogous AGL model values, and can also be different on the two spin-orbit split FSs. Since in this paper we only treat the $\uparrow$ FS, we drop the spin subscripts to simplify the notation.

The spins are quantized along ${\bm B}=B\left(\sin\theta\cos\phi,\sin\theta\sin\phi,\cos\theta\right) = {\bm\nabla}\times{\bm A} = \mu_0{\bm H}+{\bm M}$, including ${\bm M}_0$ for the ferromagnetic superconductor \cite{HH}, which we assume is non-vanishing at and below $T_c$. We neglect additional spin-orbit coupling effects that may tie the spin quantization axes to the wave vector directions, since we are only interested in parallel-spin pair states, for which the effects of spin-orbit coupling on the Zeeman energy do not significantly affect $B_{c2}$.
\section{Mean-field analytic solution of the model}
We begin with the mean-field equations of motion for the finite $T$ Green function matrix components in the presence of ${\bm B}$ \cite{SK1980}, generalized to an ellipsoidal FS,
\begin{align}
&\Bigl[i\omega_n-\sum_{j=1}^3\frac{1}{2m_j}\Bigl(\nabla_j/i-eA_j({\bm r})\Bigr)^2+\mu_{\sigma}(B)\Bigr]G_{\sigma\sigma'}({\bm r},{\bm r}',\omega_n)\nonumber\\
&\quad\quad+\sum_{\rho}\int d^3{\bm \xi}\Delta_{\sigma\rho}({\bm r},{\bm\xi})F^{\dag}_{\rho\sigma'}({\bm\xi},{\bm r}',\omega_n)=\delta_{\sigma\sigma'}\delta^3({\bm r}-{\bm r}'),\label{G}\\
&\Bigl[-i\omega_n-\sum_{j=1}^3\frac{1}{2m_j}\Bigl(i\nabla_j-eA_j({\bm r})\Bigr)^2+\mu_{\sigma}(B)\Bigr]F^{\dag}_{\sigma\sigma'}({\bm r},{\bm r}',\omega_n)\nonumber\\
&\quad\quad-\sum_{\rho}\int d^3{\bm\xi}\Delta^{*}_{\sigma\rho}({\bm r},{\bm\xi})G_{\rho\sigma'}({\bm \xi},{\bm r}',\omega_n)=0.\label{Fdagger}
\end{align}
where
\begin{align}
\Delta_{\sigma\sigma'}({\bm r},{\bm r}')&=\delta_{\sigma\sigma'}V_{\sigma\sigma}({\bm r}-{\bm r}')F_{\sigma\sigma}({\bm r},{\bm r}',0^{+})
\end{align}
is the mean-field order parameter in position and imaginary time $(\tau)$ space and the $\omega_n$ are the fermion Matsubara frequencies, the Fourier series transform variables of $\tau$.
Here and in the appendix, we have kept the spin subscripts merely to keep track of the various Green function matrix element factors for future reference, but we are presently only considering the $|\uparrow\uparrow\rangle$ spin state.

To study the full angle dependence of $B_{c2}(\theta,\phi)$, we implement the Maxwell equation-preserving Klemm-Clem (KC) transformations\cite{KC,book}, which are exact in the AGL model, and were subsequently applied to a microscopic calculation of $B_{c2}$ in \textit{d}-wave superconductors with $m_1 = m_2 < m_3$ \cite{PC}. Here we use them to calculate the effects of a general ellipsoidal FS on $B_{c2}$ for a \textit{p}-wave superconductor in the polar/CBS state, for which the order parameter anisotropy has a much stronger effect upon $B_{c2}(\theta,\phi)$ than in those $d$-wave cases\cite{PC}.
The first KC transformation is an anisotropic scale transformation that changes the ellipsoidal FS into a spherical FS \cite{book,KC}. This also changes ${\bm B}$ to ${\bm B}'=B'(\sin\theta'\cos\phi',\sin\theta'\sin\phi',\cos\theta')$, where $\theta'$ and $\phi'$ are given in the appendix. Then, one rotates $\hat{\bm B}'$ to the crystal $z'$ axis. Finally, one applies an isotropic scale transformation involving $\alpha(\theta,\phi)$ \cite{book,KC}.

After imposing gauge invariance, making use of the Helfand-Werthamer procedure based upon a Feynman theorem \cite{HW}, and Fourier transformation of the KC-transformed real-space to KC-transformed momentum-space variables, we obtain the single parallel-spin $(\uparrow\uparrow)$ linearized gap equation. The details of these calculations, including corrections of typos in the literature, are given in the appendix \cite{SK1980,HW}. We thus obtain,
\begin{align}
&\overline{\tilde{\Delta}}(\tilde{\bm R},\hat{\tilde{\bm k}})= T\sum_{\omega_{n}}\frac{N(0)}{2}\int d\Omega_{\tilde{\bm k}'}\tilde{V}(\hat{\tilde{\bm k}},\hat{\tilde{\bm k}'})\notag\\
&\quad\times\int_0^{\infty}d\xi_{\tilde{\bm k}'}e^{-2\xi_{\tilde{\bm k}'}|\omega_{n}|}e^{-i\xi_{\tilde{\bm k}'}v_F\hat{\tilde{\bm k}'}\cdot\tilde{\bm\Pi}(\tilde{\bm R})}\overline{\tilde{\Delta}}(\tilde{\bm R},\hat{\tilde{\bm k}'}),\label{gapequation}
\end{align}
where $\overline{\tilde{\Delta}}$ is the transformed $\Delta_{\uparrow\uparrow}$ amplitude without the gauge phases, $N(0)=mk_F/(2\pi^2)$ is the density of states per spin at the chemical potential $\mu_{\uparrow}(\tilde{B}_3)$ for an effectively isotropic metal with a geometric mean mass $m$, effective Fermi wave vector $k_F=\sqrt{2m\mu_{\uparrow}(\tilde{B}_3)}$, effective Fermi velocity $v_F=k_F/m$, and
\begin{align}
\tilde{\bm\Pi}(\tilde{\bm R})=-i\alpha\tilde{\bm\nabla}_{\tilde{\bm R}}-2e\tilde{\bm A}(\tilde{\bm R}),\label{Pi}
\end{align}
where $\alpha(\theta,\phi)$ is given by Eq.~(\ref{alpha}).
 We also define the anisotropy function
\begin{align}
\gamma^2(\phi)=\frac{m_3}{m_1\cos^2\phi+m_2\sin^2\phi},\label{gamma2}
\end{align}
so that $\alpha=\sqrt{\overline{m}_3}\sqrt{\cos^2\theta+\gamma^{-2}(\phi)\sin^2\theta}$. The KC transformations also modify the effective pairing interaction to become
\begin{align}
\tilde{V}(\hat{\tilde{\bm k}},\hat{\tilde{\bm k}'})=3V_0(\hat{\tilde{k}}_3\cos\theta'-\hat{\tilde{k}}_2\sin\theta')(\hat{\tilde{k}}'_3\cos\theta'-\hat{\tilde{k}}'_2\sin\theta'),\label{Vtransformed}
\end{align}
 where $\cos\theta'=\sqrt{\overline{m}_3}\cos\theta/\alpha$,
 For an isotropic ${\bm g}$ tensor, $\tilde{B}_3=B$ as the KC transformations do not modify $\mu_{\uparrow}(B)$.

The transformations have two overall effects: First, $B\rightarrow B\alpha(\theta,\phi)$ due to the transformed eigenvalues obtained from the transformed harmonic oscillator operator $\tilde{\bm\Pi}(\tilde{\bm R})$ in Eq.~(\ref{Pi}), modifying the slope of $B_{c2}$ at $T_c$ due to effective mass anisotropy, even for an $s$-wave superconductor \cite{KC,book,PC,HW,Rieck}. Second, the rotation changes $V(\hat{\bm k},\hat{\bm k}')$ to $\tilde{V}(\hat{\tilde{\bm k}},\hat{\tilde{\bm k}'})$, given by Eq.~(\ref{Vtransformed}).
 This differently alters $B_{c2}(\theta,\phi,T)$ from that of its slope at $T_c$.

 We then expand $\Delta(\tilde{\bm R},\hat{\tilde{\bm k}})$ in terms of vortex harmonic oscillator states just below ${B}_{c2}$ \cite{SK1980,SK1985},
 \begin{align}
 \Delta(\tilde{\bm R},\hat{\tilde{\bm k}})=(\hat{\tilde{k}}_3\cos\theta'-\hat{\tilde{k}}_2\sin\theta')\sum_{n=0}^{\infty}a_n|n(\tilde{\bm R})\rangle,
 \end{align}
 and obtain a general recursion relation for the expansion coefficients $a_n$,
\begin{align}
\Gamma_{n}a_{n}&=\frac{1}{2}\sin^{2}\theta'(\beta_{n}a_{n+2}+\beta_{n-2}a_{n-2}),\\
\Gamma_{n}&=-\ln t +\cos^{2}\theta'\alpha_{n}^{(p)}+\sin^{2}\theta'\alpha_{n}^{(a)},\\
\alpha_{n}^{(p,a)}&=\pi T{\sum_{\omega_{n}}}\int_{0}^{\pi}d\theta_{\tilde{\bm k}'}\sin\theta_{\tilde{\bm k}'}\left(
3\cos^{2}\theta_{\tilde{\bm k}'},
\frac{3}{2}\sin^{2}\theta_{\tilde{\bm k}'}
\right)\nonumber\\
&\quad\times\int_{0}^{\infty}d\xi_{\tilde{\bm k}'} e^{-2\xi_{\tilde{\bm k}'}|\omega_{n}|}e^{-\eta_{\tilde{\bm k}'}/2}L_{n}(\eta_{\tilde{\bm k}'}),\\
\beta_{n}&=\pi T\sum_{\omega_{n}}\int_{0}^{\pi}d\theta_{\tilde{\bm k}'}\frac{3}{2}\mathrm{sin^{3}}\theta_{\tilde{\bm k}'}\int_{0}^{\infty}d\xi_{\tilde{\bm k}'} e^{-2\xi_{\tilde{\bm k}'}|\omega_{n}|}\nonumber\\
&\quad\times e^{-\eta_{\tilde{\bm k}'}/2}(-\eta_{\tilde{\bm k}'})L_n^{(2)}(\eta_{\tilde{\bm k}'})[(n+1)(n+2)]^{-1/2},
\end{align}
where
\begin{align}
\eta_{\tilde{\bm k}'}=eB\alpha(\theta,\phi)v_{F}^{2}\xi^{2}_{\tilde{\bm k}'}\sin^{2}\theta_{\tilde{\bm k}'},
\end{align}
$t=T/T_{c}$, $T_{c}=(2e^C\omega_{0}/\pi)\exp\left(-1/N(0)V_{0}\right)$, $\omega_{0}$ is a characteristic pairing cutoff frequency, $C\approx0.5772$ is Euler's constant, and $L_{n}(z)$ and $L_n^{(2)}(z)$ are a Laguerre and an associated Laguerre polynomial, respectively.

The recursion relation for the $a_n$ differs from that obtained previously for the polar/CBS state for ${\bm B}$ in the nodal planar direction\cite{SK1985} only by the general $\theta'$ and by $B\rightarrow B\alpha(\theta,\phi)$. Solving it iteratively, $B_{c2}(\theta,\phi,t)$ is implicitly obtained from the continued-fraction equation,
\begin{align}
\Gamma_{0}-\frac{\frac{1}{4}\mathrm{sin^{4}\theta^{\prime}}\beta_{0}^{2}}{\Gamma_{2}-\frac{\frac{1}{4}\mathrm{sin^{4}\theta^{\prime}}\beta_{2}^{2}}{\Gamma_{4}\cdots}}=0.
\end{align}
Usually, $4$ or $5$ iterations yield sufficient accuracy to detect the unusual effects described in the following.

\section{Numerical results and fits to experiment}
In Fig.~2 (a), the reduced (dimensionless) magnetic induction $b_{c2}=2eB_{c2}v_{F}^{2}/(2\pi T_{c})^{2}$ is plotted versus $t$ for a spherical FS [$\gamma^2(\phi)=1$] and $\theta$ values increasing from $0^{\circ}$ [at which $b_{c2}(t)=b_{c2,{\rm p}\>{\rm antinodal}}(t)$\cite{SK1980,SK1985}] to $90^{\circ}$ [at which $b_{c2}(t)=b_{c2,{\rm planar}\>{\rm nodal}}(t)$\cite{SK1985}] from top to bottom in increments of $10^{\circ}$ \cite{SK1985}. $b_{c2}(\theta,t)$ decreases monotonically with increasing $\theta$, but is less sensitive to $\theta$ for $\theta\sim0^{\circ}$ and especially for $\theta\sim90^{\circ}$ than for ordinary FS anisotropy. As $\theta$ increases from $0^{\circ}$ to $90^{\circ}$, $-db_{c2}(\theta,t)/dt|_{t=1}$
decreases monotonically by an overall factor of $1/\sqrt{3}$. Since this slope variation is indistinguishable from that which could arise from FS anisotropy, the same curves are rescaled by $-db_{c2}/dt|_{t=1}$ in Fig.~2(b). Order parameter anisotropy effects are easiest to identify for $t\ll1$. \cite{SK1985,KS}.
\begin{figure}[!htbp]
\center{\includegraphics[width=0.48\textwidth]{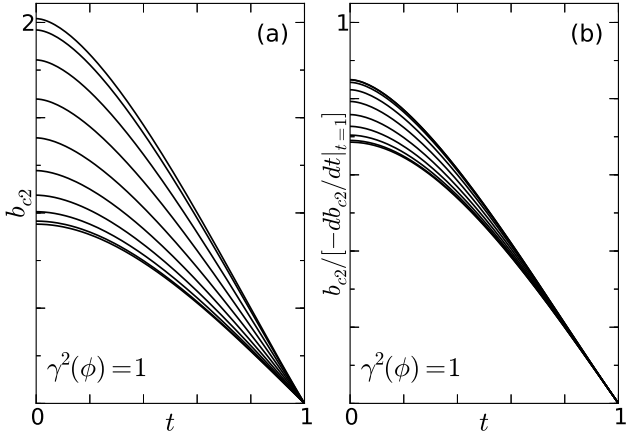}
\caption{(a) Plots of the dimensionless $b_{c2}(\theta,t)=2eB_{c2}\tilde{v}_F^2/(2\pi T_c)^2$ for the polar/CBS $p$-wave state on a spherical Fermi surface with $\theta$ increasing from $0^{\circ}$ [top, antinodal direction, with $b_{c2,{\rm p}\>{\rm antinodal}}(t)$] to 90$^{\circ}$ [bottom, planar nodal direction, with $b_{c2,{\rm planar}\>{\rm nodal}}(t)$] in increments of $10^{\circ}$. See text. (b) The same curves in Fig.~2(a) normalized by $-db_{c2}/dt|_{t=1}$.}}\label{fig2}
\end{figure}

At fixed $t$, $b_{c2}(\theta,\phi,t)$ for a polar/CBS $p$-wave superconductor with an ellipsoidal FS only depends upon $\alpha(\theta,\phi)$ and $\sin^2\theta'$, $b_{c2}(\pi-\theta,\phi,t)=b_{c2}(\theta,\phi,t)$, $\gamma^2(\phi)$ defined by Eq.~(\ref{gamma2}) contains the entire $\phi$ dependence of $b_{c2}$ \cite{book,KC},
and $-db_{c2}(\theta,\phi,t)/dt\bigr|_{t=1}\propto[3\sin^2\theta/\gamma^2(\phi)+\cos^2\theta]^{-1/2}~$\cite{SK1985}, suggesting $\gamma^2(\phi)=3$ signals a crossover from order parameter to FS anisotropy as $t\rightarrow1^{-}$.

In Fig.~3, we plotted $b_{c2}(\theta,t)/b_{c2}(0,t)$ for a variety of fixed $\gamma^2(\phi)$ values at $t=0, \frac{1}{2}$. At lower $t$ and as $\gamma^2(\phi)$ increases from 0.1 to 3, there is an increasing difference between $b_{c2}(\theta,t)$ and the effective anisotropic mass form,
\begin{align}
b_{c2}^{\rm eff}(\theta)=\left[\cos^{2}\theta/b_{c2}^{2}(0^{\circ})+\sin^{2}\theta/b_{c2}^{2}(90^{\circ})\right]^{-1/2}\label{beff} \end{align}
fitted at each $t$, which fits are indicated by the dashed curves. Anomalous peaks at $0^{\circ}<\theta^{*}<90^{\circ}$ for $\gamma^2(\phi)>3$ are indicated by the arrows. For $\gamma^2(\phi)=10$, $t=1/2$, $b_{c2}(\theta)$ only has a conventional maximum at $\theta=90^{\circ}$.
 The anomalous $b_{c2}(\theta)$ is due to competing order parameter and FS anisotropy effects.
\begin{figure}[!htbp]
\center{\includegraphics[width=0.48\textwidth]{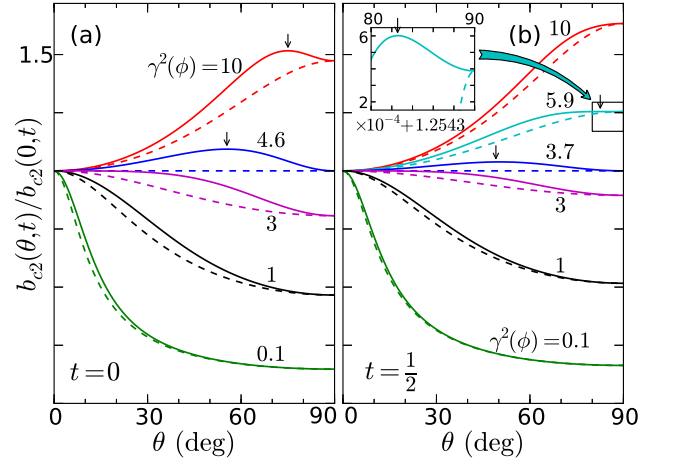}
\caption{(color online) Calculated $b_{c2}(\theta,t)/b_{c2}(0,t)$ (solid) and fitted $b_{c2}^{\rm eff}(\theta,t)/b_{c2}(0,t)$, Eq.~(\ref{beff}), (dashed) curves at constant $\gamma^2(\phi)$ values. The arrows indicate peak maxima at $\theta^{*}$ points. (a) $t=0$ (b) $t=1/2$. The inset is an enlargement of the $80^{\circ}\le\theta\le90^{\circ}$ region of the $\gamma^2(\phi)=5.9$ curve, with the indicated vertical scale points 1.2545 and 1.2549.}}\label{fig3}
\end{figure}

We extracted the FS effective masses from the RRR = 21 URhGe crystal data \cite{HH}. In Fig.~4(a) we present the calculated $b_{c2}(\theta,t)/b_{c2}(0,0)$ in the $ab$ plane (with ${\bm B}\perp\hat{\bm c}$) for different $t$ values as functions of $\theta$. The dashed lines represent fits to the corresponding fitted curves using Eq.~(\ref{beff}). Order parameter anisotropy effects in $b_{c2}(\theta)$ are significant for $t\ll1$, but not for $t \sim 1$. Since the FS anisotropy is weaker in the $ab$ plane than in the $ac$ plane, our results differ substantially in this plane from those of Eq.~(\ref{beff}). As noted above, in the $bc$ plane ($\theta=\pi/2$), $b_{c2}(\phi)\propto\gamma(\phi)$, since their $b_{c2,b}(t)$ and $b_{c2,c}(t)$ data both fit the planar nodal polar/CBS state $b_{c2,{\rm planar}\>{\rm nodal}}(t)$ \cite{HH}.

\begin{figure}[!htbp]
\center{\includegraphics[width=0.48\textwidth]{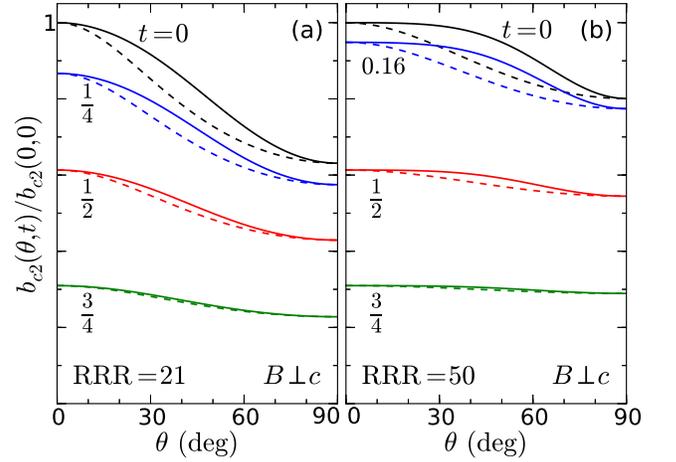}
\caption{(color online) Calculated $b_{c2}(\theta,t)/b_{c2}(0,0)$ (solid) and fitted $b_{c2}^{\rm eff}(\theta,t)/b_{c2}(0,0)$, Eq.~(\ref{beff}), (dashed) curves, for ${\bm B}\perp\hat{\bm c}$ at various $t$ values for the Fermi surface effective mass values obtained from experiment. (a) URhGe sample with RRR = 21\cite{HH}. (b) URhGe sample with RRR = 50\cite{Levy}.}}\label{fig4}
\end{figure}

We also calculated $b_{c2}(\theta,\phi,t)$ for the RRR = 50 URhGe sample \cite{Levy}. In Fig.~4(b), the calculated $b_{c2}(\theta,t)/b_{c2}(0,0)$ and correspondingly fitted curves are plotted in the $ab$ plane as a function of $\theta$ for various $t$, including $t=0.16$, the lowest measurement value \cite{Levy}. As in Fig.~4(a), the dashed curves are corresponding fits to Eq.~(\ref{beff}).

In Fig.~5, we plotted $\log_{10}[\gamma^2(\phi)]$ versus $\theta^{*}$, the anomalous peak angle in $b_{c2}(\theta,t)$. Anomalous peaks appear for $\lambda(t)>\gamma^2(\phi)>3$, where $\lambda(t)$ increases very rapidly with decreasing $t$ for $t < 0.15$, as shown in inset (a). Inset (b) details the anomalous peak in $b_{c2}(\theta,0)$ for $\gamma^2(\phi)=10^4$.

\begin{figure}[!htbp]
\center{\includegraphics[width=0.48\textwidth]{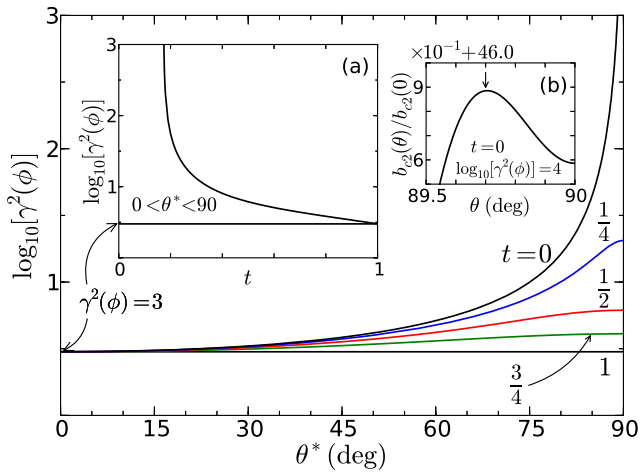}\vskip2pt
\caption{(color online) Logarithmic plot of $\gamma^2(\phi)$ as a function of $\theta^{*}$, the peak angle in $b_{c2}(\theta,t)$, at the indicated $t$ values. Inset (a): Plot of the $0^{\circ} < \theta^{*} < 90^{\circ}$ region versus $\log_{10}[\gamma^2(\phi)]$ and $t$. Inset (b): Plot of $b_{c2}(\theta,0)/b_{c2}(0,0)$ versus $\theta$ near to $\theta^{*}$ for $\gamma^2(\phi)=10^4$. The vertical scale runs from 46.5 to 47.}}\label{fig5}
\end{figure}

Conventional peaks in $b^{\rm eff}_{c2}(\theta)$ occur only at either $\theta=0^{\circ}$ or $\theta=90^{\circ}$, but anomalous peaks only occur for $0^{\circ}<\theta^{*} <90^{\circ}$. However, since $b_{c2}(\theta,\phi,t)=b_{c2}(180^{\circ}-\theta,\phi,t)$, a second anomalous peak at $180^{\circ}-\theta^{*}$ is reflection-symmetric in shape about $90^{\circ}$ to that of the first one. When $\theta^{*}$ is close to $90^{\circ}$, the magnitude of each anomalous peak is very small, but accurate measurements of this double peak could provide a definitive bulk test of the orbital symmetry of the order parameter.

\section{Discussion}
The disappearing Shubnikov de Haas (SdH) oscillations with increasing ${\bm B}$ in URhGe were claimed to be due to a topological Lifshitz FS transition and a vanishing ${\bm v}_F({\bm B})$ \cite{Yelland,Levy2}, whereas the same effect in UCoGe was claimed to be due to changes in the effective mass $m({\bm B})$ \cite{Aoki2}. Anomalously anisotropic magnetization ${\bm M}({\bm H})$ measurements of the $T$ derivative $\gamma({\bm H})$ of the specific heat in URhGe were claimed to support the latter interpretation \cite{AokiFlouquet}.
From the SdH measurements \cite{Yelland}, a strong ${\bm H}||\hat{\bm b}$ was also claimed to increase the pairing interaction strength $V_0$ and decrease the effective ${\bm v}_F({\bm B})$ \cite{Yelland} of the heavy-electron ellipsoidal FS responsible for the pairing \cite{Davis,Shick,Mueller}. We note that it could also be interpreted in terms of changes in $\{m_i({\bm B})\}$, and that $b_{c2}$ and $h_{c2}$ differ greatly for these field strengths due to the large ${\bm M}_0||\hat{\bm c}$ \cite{Levy2}. More importantly, if the order parameter in the reentrant phase maintains the polar/CBS form \cite{Mineev}, dramatic further increases in $V_0$ and potentially in $\gamma^2(\phi)$ would be expected as the metamagnetic transition is approached \cite{Yelland}, and the angle between ${\bm B}$ and ${\bm H}$ would decrease dramatically \cite{Levy2}, yielding an anomalous peak in $b_{c2}(\theta,t)$ as shown in Fig. 3. Further experiments on URhGe to measure ${\bm M}({\bm H})$ at $T_c({\bm H})$ are necessary to compare with the calculated $B_{c2}(\theta,\phi)$. Allowing $V_0\rightarrow V_0({\bm B})$ might help to fit the reentrant phase. We have calculated $\gamma({\bm B})$ self-consistently for an ellipsoidal FS in the presence of ${\bm M}_0$. These results make the analysis more complicated, but interesting. However, the derivation is lengthy, and will be published separately, along with modifications to the present fits to the URhGe $B_{c2}$ data\cite{Lorscher}.

If Sr$_2$RuO$_4$ were either a chiral (or non-chiral, depending upon the direction of ${\bm H}$) SK or a chiral ABM $p_x\pm ip_y$ parallel-spin state locked onto the layers as widely purported\cite{MM}, for ${\bm H}$ parallel to the layers $H_{c2,||}(T)$ would be proportional to either the rather linear $H_{c2,{\rm p}\>{\rm antinodal}}(T)$ or the less linear $H_{c2,{\rm ABM}\>{\rm antinodal}}(T)$\cite{Zhang,SK1980}, respectively. The former is shown as the top curve of Fig. 2(a), which differs very substantially from the experimental curves\cite{Deguchi,Kittaka,Yonezawa}, and the latter also deviates substantially\cite{Zhang}, although not as much, from the Sr$_2$RuO$_4$ parallel $H_{c2}(T)$ data that bend strongly downwards with decreasing $T$, precisely as expected for ordinary Pauli limiting \cite{book,Kittaka,Machida}, and entirely consistent with scanning tunneling microscopy results\cite{Suderow}. This is in striking contrast to $B_{c2}(T)$ measurements on URhGe and UCoGe, which violate the Pauli limit by factors of 20 or more\cite{Levy,Aoki,AokiFlouquet}, presenting very strong evidence for parallel-spin states. Fits of $H_{c2}(\theta,T)$ to different candidate Sr$_2$RuO$_4$ order parameter forms and reanalyses of the Knight shift measurements are sorely needed\cite{MM,Maeno,Hall}.

 A variational approximation to our procedure was employed to fit the similarly extremely Pauli-limited in-plane $H_{c2}(90^{\circ},\phi,t\ll1)$ of CeCu$_2$Si$_2$, in which a $d_{xy}$ order parameter was surprisingly claimed to best explain the weak ($\approx0.5$\%) azimuthal anisotropy observed\cite{Steglich}. However, that very weak azimuthal anisotropy observed in this extremely Pauli-limited situation could also be explained by a $0.5\%$ anisotropy in the $g$-tensor. Further measurements and a more accurate calculation of $H_{c2}(\theta,\phi,t)$ at intermediate $\theta$ values, where it is not dominated by Pauli-limiting effects, could provide a more definitive test of the order parameter symmetry.

 Detailed $H_{c2}(\theta,\phi,t)$ for the proposed $f$-wave forms for the C phase of UPt$_3$ could provide supporting information for that scenario\cite{MachidaMachida}. Including the intrinsic effective mass anisotropy from an ellisoidal FS of the appropriate symmetry could aid in the correct identification of the order parameter symmetry in those and many other cases. In all three of these cases, inclusion of the KC-transformed Zeeman terms with an antiparallel-spin triplet or singlet spin state would first need to be made.

\section{Summary and Conclusions}

 From analytic expressions for parallel-spin, $p$-wave superconductors with completely broken symmetry, we calculated $B_{c2}(\theta,\phi,t)$ with general ellipsoidal Fermi surface anisotropy. For fixed $m_3/(m_1\cos^2\phi+m_2\sin^2\phi)>3$, the competing effects of order parameter and Fermi surface anisotropy lead to an anomalous double peak in $B_{c2}(\theta)$ that can provide a definitive test of order parameter symmetry in URhGe and related compounds. Our method is generalizable to any order parameter symmetry, provided that the Zeeman terms are properly transformed for anti-parallel spin pairing. It is straightforward to generalize these calculations to include pairing on two spin-orbit split bands.
\begin{acknowledgments}
The authors thank J.-P.~Brison, A.~D.~Huxley, Y.~Matsuda, and K.~Scharnberg for useful discussions. This work was supported in part by
 the Florida Education Fund, the McKnight Doctoral Fellowship, a Chateaubriand Fellowship from the Embassy of France (CL), UCF startup funds (RAK), the Specialized Research Fund for the Doctoral Program of Higher Education of China (no.~20100006110021) and by Grant no.~11274039 from the National Natural Science Foundation of China.
\end{acknowledgments}

\section*{Appendix}

Here we present the details of the KC transformations on the Green functions and the resulting derivation of the microscopic gap equation\cite{KC}, and correct some typos in the literature \cite{SK1980}. Here we assume the charge of an electron is $-e$. The combined anisotropic scale transformation, rotation, and isotropic scale transformation may be written as
\begin{align}
x_{\mu}&=\frac{1}{\alpha\sqrt{\overline{m}_{\mu}}}\sum_{\nu}\lambda_{\nu\mu}\tilde{x}_{\nu},\label{position}\\
\frac{\partial}{\partial x_{\mu}}&=\alpha\sqrt{\overline{m}_{\mu}}\sum_{\nu}\lambda_{\nu\mu}\frac{\partial}{\partial\tilde{ x}_{\nu}},\label{gradient}\\
B_{\mu}&=\frac{\alpha}{\sqrt{\overline{m}_{\mu}}}\sum_{\nu}\lambda_{\nu\mu}\tilde{B}_{\nu}\nonumber\\
&=\frac{\alpha}{\sqrt{\overline{m}_{\mu}}}\lambda_{3\mu}\tilde{B}_3,\\
A_{\mu}&=\sqrt{\overline{m}_{\mu}}\sum_{\nu}\lambda_{\nu\mu}\tilde{A}_{\nu},\label{vectorpotential}
\end{align}
where
\begin{align}
\lambda=\left(\begin{array}{ccc}
\sin\phi'&-\cos\phi'&0\\
\cos\theta'\cos\phi'&\cos\theta'\sin\phi'&-\sin\theta'\\
\sin\theta'\cos\phi'&\sin\theta'\sin\phi'&\cos\theta'\end{array}\right)
\end{align}
and
\begin{align}
\sum_{\mu}\lambda_{\nu\mu}\lambda_{\nu'\mu}=\delta_{\nu\nu'}.\label{rotationidentity}
\end{align}
Note that $\lambda_{13}=0$, $\overline{m}_{\mu}=m_{\mu}/m$, $m=(m_1m_2m_3)^{1/3}$, and $\alpha=\alpha(\theta,\phi)$ is given by Eq.~(\ref{alpha}) of the text \cite{book,KC}. The transformed angles obtained after the anisotropic scale transformation are given by
\begin{align}
\cos\theta'&=\frac{\sqrt{\overline{m}_3}}{\alpha}\cos\theta,\\
\sin\theta'&=\frac{\overline{\alpha}(\phi)}{\alpha}\sin\theta,\\
\cos\phi'&=\frac{\sqrt{\overline{m}_1}\cos\phi}{\overline{\alpha}(\phi)},\\
\sin\phi'&=\frac{\sqrt{\overline{m}_2}\sin\phi}{\overline{\alpha}(\phi)},\\
\overline{\alpha}(\phi)&=\alpha(\pi/2,\phi)=[\overline{m}_1\cos^2\phi+\overline{m}_2\sin^2\phi]^{1/2}\nonumber\\
&=\frac{\overline{m}_3}{\gamma^2(\phi)}
\end{align}
We begin with Eqs.~(\ref{G})-(\ref{Fdagger}) of the text. To transform the quadratic operators on the left-hand sides, we expand the gradient and vector potential components with Eqs.~(\ref{gradient}) and (\ref{vectorpotential}), and make use of the rotation identity, Eq.~(\ref{rotationidentity}). With regard to the delta function in Eq.~(\ref{G}), it is easily seen that
\begin{align}
\delta^3({\bm r})&=\int\frac{d^3{\bm k}}{(2\pi)^3}e^{i{\bm k}\cdot{\bm r}}\notag\\
&\rightarrow\alpha^3\int\frac{d^3\tilde{\bm k}}{(2\pi)^3}e^{i\tilde{\bm k}\cdot\tilde{\bm r}}
=\alpha^3\delta^3(\tilde{\bm r}).\label{deltafunction}
\end{align}
We note that $d^3{\bm k}\rightarrow\alpha^3(\overline{m}_1\overline{m}_2\overline{m}_3)^{1/2}d^3\tilde{\bm k}=\alpha^3d^3\tilde{\bm k}$,
 as the transformed volume element is invariant under all rotations. Note that to transform ${\bm k}\cdot{\bm r}$ in the exponent, expand the components of ${\bm r}$ and ${\bm k}$ according to Eqs.~(\ref{position}) and (\ref{gradient}), and again make use of the rotation identity, Eq.~(\ref{rotationidentity}). Note that the scalar product of two vectors is invariant under all rotations.

We then may write the transformed Eqs.~(\ref{G}) and (\ref{Fdagger}) of the text as
\begin{align*}
&\Bigl[i\omega_n-\frac{1}{2\tilde{m}}\Bigl(\tilde{\bm\nabla}/i-\tilde{e}\tilde{\bm A}(\tilde{\bm r})\Bigr)^2+\mu_{\sigma}(\tilde{B}_3)\Bigr]\tilde{G}_{\sigma\sigma'}(\tilde{\bm r},\tilde{\bm r}',\omega_n)\nonumber\\
&\quad+\sum_{\rho}\int d^3\tilde{\bm \xi}\tilde{\Delta}_{\sigma\rho}(\tilde{\bm r},\tilde{\bm\xi})\tilde{F}^{\dag}_{\rho\sigma'}(\tilde{\bm\xi},\tilde{\bm r}',\omega_n)=\alpha^3\delta_{\sigma\sigma'}\delta^3(\tilde{\bm r}-\tilde{\bm r}'),\\
&\Bigl[-i\omega_n-\frac{1}{2\tilde{m}}\Bigl(i\tilde{\bm\nabla}-\tilde{e}\tilde{\bm A}(\tilde{\bm r})\Bigr)^2+\mu_{\sigma}(\tilde{B}_3)\Bigr]\tilde{F}^{\dag}_{\sigma\sigma'}(\tilde{\bm r},\tilde{\bm r}',\omega_n)\nonumber\\
&\quad-\sum_{\rho}\int d^3\tilde{\bm\xi}\tilde{\Delta}^{*}_{\sigma\rho}(\tilde{\bm r},\tilde{\bm\xi})\tilde{G}_{\rho\sigma'}(\tilde{\bm \xi},\tilde{\bm r}',\omega_n)=0,
\end{align*}
where
\begin{align}
\tilde{e}&=e/\alpha,\\
\tilde{m}&=m/\alpha^2\label{tildem}
\end{align}
are the renormalized electronic charge magnitude and mass due to the transformations, and
 $\tilde{G}$, $\tilde{F}^{\dag}$ and $\tilde{\Delta}$ are complicated functions of the transformed variables, since the interaction is best determined in momentum space, as in Eq.~(\ref{interaction}).

Now in order to make the transformed functions gauge invariant, we require the equations of motion in the variables $\tilde{\bm r}$ and $\tilde{\bm r}'$ to be respectively invariant under
\begin{align}
\tilde{\bm A}(\tilde{\bm r})&=\tilde{\bm A}_0(\tilde{\bm r})+\tilde{\bm\nabla}\Phi(\tilde{\bm r}),\\
\tilde{\bm A}(\tilde{\bm r}')&=\tilde{\bm A}_0(\tilde{\bm r}')+\tilde{\bm\nabla}'\Phi(\tilde{\bm r}'),
\end{align}
where $\tilde{A}_0$ can be taken to vanish.
We then may write
\begin{align}
\tilde{G}_{\sigma\sigma'}(\tilde{\bm r},\tilde{\bm r}',\omega_n)&=\overline{\tilde{G}}_{\sigma\sigma'}(\tilde{\bm r},\tilde{\bm r}',\omega_n)e^{i\tilde{e}[\Phi(\tilde{\bm r})-\Phi(\tilde{\bm r}')]},\label{Gplusbar}\\
\tilde{F}^{\dag}_{\sigma\sigma'}(\tilde{\bm r},\tilde{\bm r}',\omega_n)&=\overline{\tilde{F}}^{\dag}_{\sigma\sigma'}(\tilde{\bm r},\tilde{\bm r}',\omega_n)e^{-i\tilde{e}[\Phi(\tilde{\bm r})+\Phi(\tilde{\bm r}')]},\label{Fdaggerbar}\\
\tilde{G}_{\sigma\sigma'}(\tilde{\bm r},\tilde{\bm r}',-\omega_n)&=\overline{\tilde{G}}_{\sigma\sigma'}(\tilde{\bm r},\tilde{\bm r}',-\omega_n)e^{i\tilde{e}[\Phi(\tilde{\bm r}')-\Phi(\tilde{\bm r})]},\label{Gminusbar}\\
\tilde{F}_{\sigma\sigma'}(\tilde{\bm r},\tilde{\bm r}',\omega_n)&=\overline{\tilde{F}}_{\sigma\sigma'}(\tilde{\bm r},\tilde{\bm r}',\omega_n)e^{i\tilde{e}[\Phi(\tilde{\bm r})+\Phi(\tilde{\bm r}')]},\label{Fbar}
\end{align}
as was done long ago for isotropic superconductors\cite{AGD}.

We then examine the bare Green functions in the absence of any pairing. We have
\begin{align}
\Bigl[\pm i\omega_n-&\frac{1}{2\tilde{m}}\Bigl(\tilde{\bm\nabla}/i\mp \tilde{e}\tilde{\bm A}(\tilde{\bm r})\Bigr)^2+\mu_{\sigma}(\tilde{B}_3)\Bigr]\notag\\
&\times\tilde{G}^0_{\sigma\sigma'}(\tilde{\bm r},\tilde{\bm r}',\pm\omega_n)=\delta_{\sigma\sigma'}\alpha^3\delta(\tilde{\bm r}-\tilde{\bm r}').
\end{align}
These forms are easily shown to satisfy
\begin{align}
\tilde{G}^0_{\sigma\sigma'}(\tilde{\bm r},\tilde{\bm r}',\pm\omega_n)=\overline{\tilde{G}}^0_{\sigma\sigma'}(\tilde{\bm r},\tilde{\bm r}',\pm\omega_n)e^{\mp i\tilde{e}\phi(\tilde{\bm r},\tilde{\bm r}')},
\end{align}
where
\begin{align}
\phi(\tilde{\bm r},\tilde{\bm r}')=\int_{\tilde{\bm r}}^{\tilde{\bm r}'}\tilde{A}(\tilde{\bm s})\cdot d\tilde{\bm s}=\Phi(\tilde{\bm r}')-\Phi(\tilde{\bm r}),
\end{align}
precisely as for a spherical Fermi surface, except that \mbox{$e\rightarrow\tilde{e}$} and \mbox{$m\rightarrow\tilde{m}$}. We note that $\overline{\tilde{G}}^0_{\sigma\sigma'}(\tilde{\bm r},\tilde{\bm r}',\pm\omega_n)$ satisfies
\begin{align}
 \Bigl[\pm i\omega_n+\frac{\tilde{\bm\nabla}^2}{2\tilde{m}}+\mu_{\sigma}(\tilde{B}_3)\Bigr]\overline{\tilde{G}}^0_{\sigma\sigma'}&(\tilde{\bm r},\tilde{\bm r}',\pm\omega_n)\notag\\
 &=\delta_{\sigma\sigma'}\alpha^3\delta^3(\tilde{\bm r}-\tilde{\bm r}'),\label{G0}
 \end{align}
 which can be taken to be a function of $\tilde{\bm r}-\tilde{\bm r}'$, and can therefore be Fourier transformed. Writing
 \begin{align*}
 \delta^3(\tilde{\bm r}-\tilde{\bm r}')&=\int\frac{d^3\tilde{\bm k}}{(2\pi)^3}e^{i\tilde{\bm k}\cdot(\tilde{\bm r}-\tilde{\bm r}')},\\
 \overline{\tilde{G}}^0_{\sigma\sigma'}(\tilde{\bm r}-\tilde{\bm r}',\pm\omega_n)&=\alpha^3\int\frac{d^3\tilde{\bm k}}{(2\pi)^3}\overline{\tilde{G}}^0_{\sigma\sigma'}(\tilde{\bm k},\pm\omega_n)e^{i\tilde{\bm k}\cdot(\tilde{\bm r}-\tilde{\bm r}')},
 \end{align*}
 and using Eq.~(\ref{G0}), we easily obtain
 \begin{align*}
 \overline{\tilde{G}}^0_{\sigma\sigma'}(\tilde{\bm k},\pm\omega_n)=\frac{\delta_{\sigma\sigma'}}{\pm i\omega_n-\tilde{\bm k}^2/(2\tilde{m})+\mu_{\sigma}(\tilde{B}_3)}.
 \end{align*}
 To obtain $\overline{\tilde{G}}^0_{\sigma\sigma'}(\tilde{\bm r},\tilde{\bm r}',\pm\omega_n)$ in real space, one can easily perform the same contour integral as was done long ago for isotropic superconductors on a spherical FS\cite{AGD}, obtaining
 \begin{align}
 &\quad\,\overline{\tilde{G}}^0_{\sigma\sigma'}(\tilde{\bm r}-\tilde{\bm r}',\pm\omega_n)\notag\\
 &=\alpha^3\int\frac{d^3\tilde{\bm k}}{(2\pi)^3}e^{i\tilde{\bm k}\cdot(\tilde{\bm r}-\tilde{\bm r}')}\frac{\delta_{\sigma\sigma'}}{\pm i\omega_n-\frac{\tilde{\bm k}^2}{2\tilde{m}}+\mu_{\sigma}(\tilde{B}_3)}\label{G0integral}\\
 &=-\frac{\delta_{\sigma\sigma'}\tilde{m}\alpha^3}{2\pi|\tilde{\bm r}-\tilde{\bm r}'|}e^{\pm i\tilde{k}_F|\tilde{\bm r}-\tilde{\bm r}'|{\rm sgn}(\omega_n)-|\omega_n||\tilde{\bm r}-\tilde{\bm r}'|/\tilde{v}_F},\label{realspaceG0}
 \end{align}
 where $\tilde{k}_F=k_F/\alpha$, $\tilde{v}_F=\alpha v_F$, and $\tilde{m}$ is given by Eq.~(\ref{tildem}). In deriving Eq.~(\ref{realspaceG0}), it is easiest to first perform the angular integrals, and then to note that
\begin{align}
 &\quad\,\int_0^{\infty}\tilde{k}d\tilde{k}(e^{i\tilde{k}|\tilde{\bm r}-\tilde{\bm r}'|}-e^{-i\tilde{k}|\tilde{\bm r}-\tilde{\bm r}'|})\overline{\tilde{G}}^0_{\sigma\sigma'}(\tilde{\bm k},\pm\omega_m)\notag\\
 &=\int_{-\infty}^{\infty}\tilde{k}d\tilde{k}e^{i\tilde{k}|\tilde{\bm r}-\tilde{\bm r}'|}\overline{\tilde{G}}^0_{\sigma\sigma'}(\tilde{\bm k},\pm\omega_m)\label{Iplus}\\
 &=-\int_{-\infty}^{\infty}\tilde{k}d\tilde{k}e^{-i\tilde{k}|\tilde{\bm r}-\tilde{\bm r}'|}\overline{\tilde{G}}^0_{\sigma\sigma'}(\tilde{\bm k},\pm\omega_m).\label{Iminus}
\end{align}
 Then set $\tilde{k}=\tilde{k}_F+{\tilde k}-\tilde{k}_F$, let $\mu_{\sigma}(\tilde{B}_3)=\tilde{k}_F^2/(2\tilde{m})$, set $\xi_{\tilde k}=(\tilde{k}-\tilde{k}_F)\tilde{v}_F$, and neglect the term proportional to $\xi^2_{\tilde{k}}$. Then, if $\pm\omega_n>0$, use Eq.~(\ref{Iplus}), and close the contour in the upper half plane. If $\pm\omega_n<0$, use Eq.~(\ref{Iminus}), and close the contour in the lower half plane. Note that the sum over the $\omega_n$ is performed after the final gap equation is evaluated, so there is a single pole at $\xi_{{k}}=\pm i\omega_n$ in Eq.~(\ref{G0integral}).

 Equations (\ref{G}) and (\ref{Fdagger}) may be rewritten as
 \begin{align}
 \tilde{G}_{\sigma\sigma'}(\tilde{\bm r},\tilde{\bm r}',\omega_n)&=\tilde{G}^0_{\sigma\sigma}(\tilde{\bm r},\tilde{\bm r}',\omega_n)\delta_{\sigma\sigma'}\notag\\
&\quad-\alpha^{-6}\int d^3\tilde{\bm\xi}d^3\tilde{\bm \xi}'\sum_{\rho}\tilde{G}^0_{\sigma\sigma}(\tilde{\bm r},\tilde{\bm \xi}',\omega_n)\notag\\
&\qquad\times\tilde{\Delta}_{\sigma\rho}(\tilde{\bm\xi}',\tilde{\bm\xi})\tilde{F}^{\dag}_{\rho\sigma'}(\tilde{\bm\xi},\tilde{\bm r}',\omega_n),\\
\tilde{F}_{\sigma\sigma'}(\tilde{\bm r},\tilde{\bm r}',\omega_n)&=\alpha^{-6}\int d^3\tilde{\bm\xi}d^3\tilde{\bm\xi}'\sum_{\rho}
\tilde{G}^0_{\sigma\sigma}(\tilde{\bm r},\tilde{\bm\xi},\omega_n)\notag\\
&\qquad\times\tilde{\Delta}_{\sigma\rho}(\tilde{\bm\xi},\tilde{\bm\xi}')\tilde{G}_{\rho\sigma'}(\tilde{\bm r}',\tilde{\bm\xi}',-\omega_n),
 \end{align}
 where the two factors of $\alpha^{-3}$ arise from the KC transformations, since
 the volume element is rotationally invariant, and hence $d^3{\bm \xi}\rightarrow\alpha^{-3}(\overline{m}_1\overline{m}_2\overline{m}_3)^{-1/2}d^3\tilde{\bm\xi}=\alpha^{-3}d^3\tilde{\bm \xi}$.

 In real space and imaginary time, the superconducting order parameter is defined by
 \begin{align}
 \tilde{\Delta}_{\sigma\sigma'}(\tilde{\bm r},\tilde{\bm r}')=\tilde{V}(\tilde{\bm r}-\tilde{\bm r}')\tilde{F}_{\sigma\sigma'}(\tilde{\bm r},\tilde{\bm r}',0^{+}),
 \end{align}
 resulting in the gap equation in the transformed variables,
 \begin{align}
 \tilde{\Delta}_{\sigma\sigma'}&(\tilde{\bm r},\tilde{\bm r}')=\tilde{V}(\tilde{\bm r}-\tilde{\bm r}')\alpha^{-6}T\sum_{\omega_n}\sum_{\rho}\int d^3\tilde{\bm\xi}d^3\tilde{\bm\xi}'\notag\\
&\times\tilde{G}^0_{\sigma\sigma}(\tilde{\bm r},\tilde{\bm\xi},\omega_n)\tilde{\Delta}_{\sigma\rho}(\tilde{\bm\xi},\tilde{\bm\xi}')\tilde{G}_{\rho\sigma'}(\tilde{\bm r}',\tilde{\bm\xi}',-\omega_n).
\end{align}
Since the order parameter is obtained from the $\tilde{F}$ function, we have to include it to insure gauge invariance. Thus, we write
\begin{align}
\tilde{\Delta}_{\sigma\sigma'}(\tilde{\bm r},\tilde{\bm r}')=\overline{\tilde{\Delta}}_{\sigma\sigma'}(\tilde{\bm r},\tilde{\bm r}')e^{i\tilde{e}[\Phi(\tilde{\rm r})+\Phi(\tilde{\rm r}')]}.\label{Deltabar}
\end{align}
Using Eqs.~(\ref{Gplusbar}), (\ref{Gminusbar}), and (\ref{Deltabar}), and after dividing by the exponents in Eq.~(\ref{Deltabar}), we obtain
\begin{align*}
&\overline{\tilde{\Delta}}_{\sigma\sigma'}(\tilde{\bm r},\tilde{\bm r}')=\tilde{V}(\tilde{\bm r}-\tilde{\bm r}')\alpha^{-6}T\sum_{\omega_n}\sum_{\rho}\int d^3\tilde{\bm\xi}d^3\tilde{\bm\xi}'\notag\\
&\quad\times\overline{\tilde{G}}^0_{\sigma\sigma}(\tilde{\bm r}-\tilde{\bm \xi},\omega_n)e^{2i\tilde{e}\phi(\tilde{\bm r}',\tilde{\bm \xi}')}\overline{\tilde{\Delta}}_{\sigma\rho}(\tilde{\bm\xi},\tilde{\bm\xi}')\overline{\tilde{G}}_{\rho\sigma'}(\tilde{\bm\xi}',\tilde{\bm r}',-\omega_n).
\end{align*}
We then rewrite the order parameter and the full Green function in terms of their centers of mass and relative positions, obtaining
\begin{align*}
&\overline{\tilde{\Delta}}_{\sigma\sigma'}\bigl(\frac{\tilde{\bm r}+\tilde{\bm r}'}{2},\tilde{\bm r}-\tilde{\bm r}'\bigr)=\tilde{V}(\tilde{\bm r}-\tilde{\bm r}')\alpha^{-6}T\notag\\
&\quad\times\sum_{\omega_n}\sum_{\rho}\int d^3\tilde{\bm\xi}d^3\tilde{\bm\xi}'
\overline{\tilde{G}}^0_{\sigma\sigma}(\tilde{\bm r}-\tilde{\bm \xi},\omega_n)
e^{2i\tilde{e}\int_{\tilde{\bm r}'}^{\tilde{\bm\xi}'}\tilde{\bm A}(\tilde{\bm s})\cdot d\tilde{\bm s}}\nonumber\\
&\qquad\times\overline{\tilde{\Delta}}_{\sigma\rho}\bigl(\frac{\tilde{\bm\xi}+\tilde{\bm\xi}'}{2},\tilde{\bm\xi}-\tilde{\bm\xi}'\bigr)
\overline{\tilde{G}}_{\rho\sigma'}
\bigl(\frac{\tilde{\bm r}'+\tilde{\bm\xi}'}{2},\tilde{\bm r}'-\tilde{\bm\xi}',-\omega_n\bigr).
\end{align*}
Now, we let
\begin{align}
\tilde{\bm R}=\frac{\tilde{\bm r}+\tilde{\bm r}'}{2}
\end{align}
be the center of mass of the unperturbed order parameter. Thus, we may rewrite
\begin{align*}
&\overline{\tilde{\Delta}}_{\sigma\rho}\bigl(\frac{\tilde{\bm\xi}+\tilde{\bm\xi}'}{2},\tilde{\bm\xi}-\tilde{\bm\xi}'\bigr)\notag\\
&\qquad=e^{\bigl(\frac{\tilde{\bm\xi}+\tilde{\bm\xi}'}{2}-\tilde{\bm r}''\bigr)\cdot\tilde{\bm\nabla}_{\tilde{\bm R}}}\overline{\tilde{\Delta}}_{\sigma\rho}(\tilde{\bm R},\tilde{\bm\xi}-\tilde{\bm\xi}')\Bigr|_{\tilde{\bm r}''=\tilde{\bm R}},\\
&\overline{\tilde{G}}_{\rho\sigma'}\bigl(\frac{\tilde{\bm r}'+\tilde{\bm\xi}'}{2},\tilde{\bm r}'-\tilde{\bm\xi}',-\omega_n\bigr)\notag\\
&\quad=e^{\bigl(\frac{\tilde{\bm r}'+\tilde{\bm\xi}'}{2}-\tilde{\bm r}'''\bigr)\cdot\tilde{\bm\nabla}_{\tilde{\bm R}}}\overline{\tilde{G}}_{\rho\sigma'}(\tilde{\bm R},\tilde{\bm r}'-\tilde{\bm\xi}',-\omega_n)\Bigr|_{\tilde{\bm r}'''=\tilde{\bm R}}.
\end{align*}
Note that these operations are just reformulations of the Taylor series expansions.

We then make the approximations that $\tilde{\bm R}=(\tilde{\bm r}+\tilde{\bm r}')/2\approx\tilde{\bm r}'$ and $(\tilde{\bm\xi}+\tilde{\bm\xi}')/2\approx\tilde{\bm\xi}'$, as the center of mass of the order parameter is close to the positions of either paired electron. Then
\begin{align*}
&e^{2i\tilde{e}\phi(\tilde{\bm r}',\tilde{\bm \xi}')}\, \overline{\tilde{\Delta}}\bigl(\frac{\tilde{\bm\xi}+\tilde{\bm\xi}'}{2},\tilde{\bm\xi}-\tilde{\bm\xi}'\bigr)\notag\\
&\qquad\qquad\qquad\times\overline{\tilde{G}}_{\rho\sigma'}\bigl(\frac{\tilde{\bm r}'+\tilde{\bm \xi}'}{2},\tilde{\bm r}'-\tilde{\bm\xi}',-\omega_n\bigr)\notag\\
\approx\, &e^{2i\tilde{e}\phi(\tilde{\bm r}',\tilde{\bm \xi}')}\,e^{(\tilde{\bm\xi}'-\tilde{\bm r}')\cdot\tilde{\bm\nabla}_{\tilde{\bm R}}}\,\overline{\tilde{\Delta}}_{\sigma\rho}(\tilde{\bm R},\tilde{\bm\xi}-\tilde{\bm\xi}')\notag\\
&\qquad\qquad\times e^{\frac{1}{2}( \tilde{\bm\xi}'-\tilde{\bm r}')\cdot\tilde{\bm\nabla}_{\tilde{\bm R}}}\,\overline{\tilde{G}}_{\rho\sigma'}(\tilde{\bm R},\tilde{\bm r}'-\tilde{\bm\xi}',-\omega_n)\notag\\
=\, &e^{(\tilde{\bm\xi}'-\tilde{\bm r}')\cdot[\tilde{\bm\nabla}_{\tilde{\bm R}}-2i\tilde{e}\tilde{\bm A}(\tilde{\bm R})]}\,\overline{\tilde{\Delta}}_{\sigma\rho}(\tilde{\bm R},\tilde{\bm\xi}-\tilde{\bm\xi}')\notag\\
&\qquad\qquad\times e^{\frac{1}{2}( \tilde{\bm\xi}'-\tilde{\bm r}')\cdot\tilde{\bm\nabla}_{\tilde{\bm R}}}\,\overline{\tilde{G}}_{\rho\sigma'}(\tilde{\bm R},\tilde{\bm r}'-\tilde{\bm\xi}',-\omega_n),
\end{align*}
where we set $\tilde{\bm r}'\approx\tilde{\bm R}$ and $(\tilde{\bm\xi}+\tilde{\bm \xi}')/2\approx\tilde{\bm\xi}'$, and made use of the Helfand-Werthamer procedure based upon a Feynman theorem\cite{HW}. Thus, the gap equation may be written as
\begin{align*}
&\overline{\tilde{\Delta}}_{\sigma\sigma'}(\tilde{\bm R},\tilde{\bm r}-\tilde{\bm r}')=
\tilde{V}(\tilde{\bm r}-\tilde{\bm r}')\alpha^{-6}T\notag\\
&\quad\times\sum_{\omega_n}\sum_{\rho}\int d^3\tilde{\bm\xi}d^3\tilde{\bm\xi}'\,\overline{\tilde{G}}^0_{\sigma\sigma}(\tilde{\bm r}-\tilde{\bm \xi},\omega_n)\,e^{i(\tilde{\bm\xi}'-\tilde{\bm r}')\cdot\tilde{\bm{\Pi}}(\tilde{\bm R})/\alpha}\notag\\
&\qquad\times\overline{\tilde{\Delta}}_{\sigma\rho}(\tilde{\bm R},\tilde{\bm\xi}-\tilde{\bm\xi}')\,e^{\frac{1}{2}(\tilde{\bm\xi}'-\tilde{\bm r}')\cdot\tilde{\bm\nabla}_{\tilde{\bm R}}}\,\overline{\tilde{G}}_{\rho\sigma'}(\tilde{\bm R},\tilde{\bm r}'-\tilde{\bm\xi}',-\omega_n),
\end{align*}
where $\tilde{\bm\Pi}(\tilde{\bm R})$ is given by Eq.~(\ref{Pi}) of the text. We note that this expression differs slightly from that obtained previously, due to an unfortunate typo that interchanged $G$ with $G^0$\cite{SK1980}. To clarify that this result is correct, we put in the spin indices to preserve the matrix multiplications correctly. This change does not affect the behavior at $H_{c2}$, however.

We note that at (or just barely below) $H_{c2}$ (or $B_{c2}$), the order parameter is vanishingly small, so it suffices to set
\begin{align}
\overline{\tilde{G}}_{\rho\sigma'}(\tilde{\bm R},\tilde{\bm r}'-\tilde{\bm\xi}',-\omega_n)\approx\overline{\tilde{G}}^0_{\rho\sigma'}(\tilde{\bm r}'-\tilde{\bm \xi}',-\omega_n),
\end{align}
 which is independent of $\tilde{\bm R}$, and hence the factor $e^{\frac{1}{2}(\tilde{\bm\xi}'-\tilde{\bm r}')\cdot\tilde{\bm\nabla}_{\tilde{\bm R}}}$ can be set equal to unity. We thus have the equation in real space for the calculation of $B_{c2}$,
\begin{align}
\overline{\tilde{\Delta}}_{\sigma\sigma'}&(\tilde{\bm R},\tilde{\bm r}-\tilde{\bm r}')=
\tilde{V}(\tilde{\bm r}-\tilde{\bm r}')\alpha^{-6}T\notag\\
&\times\sum_{\omega_n}\int d^3\tilde{\bm\xi}d^3\tilde{\bm\xi}'\,\overline{\tilde{G}}^0_{\sigma\sigma}(\tilde{\bm r}-\tilde{\bm \xi},\omega_n)e^{i(\tilde{\bm\xi}'-\tilde{\bm r}')\cdot\tilde{\bm{\Pi}}(\tilde{\bm R})/\alpha}\notag\\
&\qquad\times\overline{\tilde{\Delta}}_{\sigma\sigma'}(\tilde{\bm R},\tilde{\bm\xi}-\tilde{\bm\xi}')\overline{\tilde{G}}^0_{\sigma'\sigma'}(\tilde{\bm r}'-\tilde{\bm\xi}',-\omega_n).\label{gapequationrealspace}
\end{align}
We remark that the pairing interaction is best defined in momentum space, so we have to transform this equation to the KC-transformed momentum space, which will allow us to properly transform the pairing interaction. Hence, we shall include enough intermediate steps to demonstrate the correct $\alpha$ dependence of the KC-transformed gap equation.

In order to Fourier transform the right-hand side of Eq.~(\ref{gapequationrealspace}), we first let $\tilde{\bm \xi}\rightarrow\tilde{\bm\xi}+\tilde{\bm r}$ and $\tilde{\bm \xi}'\rightarrow\tilde{\bm\xi}'+\tilde{\bm r}'$. This means we only need to Fourier transform $\overline{\tilde{\Delta}}_{\sigma\rho}(\tilde{\bm R},\tilde{\bm\xi}+\tilde{\bm r}-\tilde{\bm\xi}'-\tilde{\bm r}')$ to obtain all of the $\tilde{\bm r}-\tilde{\bm r}'$ terms in the exponent for comparison with that in Eq.~(\ref{DeltaFT}). In writing the Fourier transform, we use the same transformation $d^3{\bm k}\rightarrow\alpha^3d^3\tilde{\bm k}$ as in Eq.~(\ref{deltafunction}). We then obtain
\begin{align}
 \overline{\tilde{\Delta}}_{\sigma\sigma'}(\tilde{\bm R},\tilde{\bm k})&=\alpha^{-3}\int\frac{d^3\tilde{\bm k}'}{(2\pi)^3}e^{i\tilde{\bm k}'\cdot(\tilde{\bm \xi}-\tilde{\bm\xi}')}T\sum_{\omega_n}\tilde{V}(\tilde{\bm k}-\tilde{\bm k}')\nonumber\\
 &\quad\times\int d^3\tilde{\bm \xi}d^3\tilde{\bm \xi}'\overline{\tilde{G}}^0_{\sigma\sigma}(\tilde{\bm\xi}',\omega_n)
 e^{i\tilde{\bm\xi}\cdot\tilde{\bm\Pi}(\tilde{\bm R})/\alpha}\notag\\
 &\qquad\quad\times\overline{\tilde{\Delta}}_{\sigma\sigma'}(\tilde{\bm R},\tilde{\bm k}')\overline{\tilde{G}}^0_{\sigma'\sigma'}(\tilde{\bm \xi},-\omega_n),\label{DeltaFT}
\end{align}
where we interchanged $\tilde{\bm\xi}$ and $\tilde{\bm\xi}'$ for convenience, and we assumed the sample to exhibit inversion symmetry in the absence of a magnetic field.

We now need to write the transformed interaction $\tilde{V}(\tilde{\bm k}-\tilde{\bm k}')$ explicitly. We first note that the relevant part of an untransformed interaction of the form
$V_0[(\hat{\bm k}-\hat{\bm k}')^2-2]=-2V_0\hat{\bm k}\cdot\hat{\bm k}'$,
is rotationally invariant, as studied previously \cite{SK1980}. However, if we break this symmetry, and only allow the pairing to be in one or two dimensions, we could have the relevant bare interaction be as described in the text,
$V(\hat{\bm k},\hat{\bm k}')=V_0\hat{k}_3\hat{k}_3'$,
where $\hat{k}_3$ is given by Eq.~(\ref{kihat}) with $i=3$. Then, making the KC transformations, we obtain
\begin{align}
\tilde{V}(\hat{\tilde{\bm k}},\hat{\tilde{\bm k}}')=3V_0(\hat{\tilde{k}}_3\cos\theta'-\hat{\tilde{k}}_2\sin\theta')(\hat{\tilde{k}}'_3\cos\theta'-\hat{\tilde{k}}'_2\sin\theta'),\label{tildeV}
\end{align}
This leads to
\begin{align}
\overline{\tilde{\Delta}}_{\sigma\sigma'}&(\tilde{\bm R},\tilde{\bm k})=T\alpha^{-3}\sum_{\omega_n}\int\frac{d^3\tilde{\bm k}'}{(2\pi)^3}e^{i\tilde{\bm k}'\cdot(\tilde{\bm \xi}-\tilde{\bm\xi}')}\nonumber\\
 &\times 3V_0(\hat{\tilde{k}}_3\cos\theta'-\hat{\tilde{k}}_2\sin\theta')(\hat{\tilde{k}}_3'\cos\theta'-\hat{\tilde{k}}'_2\sin\theta')\nonumber\\
 &\quad\times\int d^3\tilde{\bm \xi}d^3\tilde{\bm \xi}'\overline{\tilde{G}}^0_{\sigma\sigma}(\tilde{\bm\xi}',\omega_n)
 e^{i\tilde{\bm\xi}\cdot\tilde{\bm\Pi}(\tilde{\bm R})/\alpha}\overline{\tilde{\Delta}}_{\sigma\sigma'}(\tilde{\bm R},\tilde{\bm k}')\notag\\
 &\qquad\times\overline{\tilde{G}}^0_{\sigma'\sigma'}(\tilde{\bm \xi},-\omega_n),
\end{align}
We then may write
\begin{align}
\overline{\tilde{\Delta}}_{\sigma\sigma'}(\tilde{\bm R},\tilde{\bm k})=\overline{\tilde{\Delta}}_{\sigma\sigma'}(\tilde{\bm R})(\hat{\tilde{k}}_3\cos\theta'-\hat{\tilde{k}}_2\sin\theta'),
\end{align}
leading to
\begin{align}
\overline{\tilde{\Delta}}_{\sigma\sigma'}(\tilde{\bm R})&=T\alpha^{-3}\sum_{\omega_n}3V_0\int\frac{d^3\tilde{\bm k}'}{(2\pi)^3}e^{i\tilde{\bm k}'\cdot(\tilde{\bm\xi}-\tilde{\bm\xi}')}\notag\\
&\times(\hat{\tilde{k}}'_3\cos\theta'-\hat{\tilde{k}}'_2\sin\theta')^2\int d^3\tilde{\bm\xi}d^3\tilde{\bm\xi}'\overline{\tilde{G}}^0_{\sigma\sigma}(\tilde{\bm\xi}',\omega_n)\notag\\
&\quad\times e^{i\tilde{\bm\xi}\cdot\tilde{\bm\Pi}(\tilde{\bm R})/\alpha}\overline{\tilde{\Delta}}_{\sigma\sigma'}(\tilde{\bm R})\overline{\tilde{G}}^0_{\sigma'\sigma'}(\tilde{\bm\xi},-\omega_n).
\end{align}

Then, we invoke the mild approximation used previously\cite{SK1980},
\begin{align}
\int d^3\tilde{\bm k}'e^{i\tilde{\bm k}'\cdot(\tilde{\bm\xi}-\tilde{\bm\xi}')}\hat{\tilde{k}}'_{\mu}\hat{\tilde{k}}'_{\nu}=(2\pi)^3\hat{\tilde{\xi}}_{\mu}\hat{\tilde{\xi}}_{\nu}\delta^3(\tilde{\bm\xi}-\tilde{\bm\xi}'),
\end{align}
which also works with the transformed variables.
This leads to
\begin{align}
\overline{\tilde{\Delta}}_{\sigma\sigma'}(\tilde{\bm R})&=T\alpha^{-3}3V_0\sum_{\omega_n}\int d^3\tilde{\bm\xi}'(\hat{\tilde{\xi}}'_3\cos\theta'-\hat{\tilde{\xi}}'_2\sin\theta')^2\nonumber\\
\times \overline{\tilde{G}}^0_{\sigma\sigma}(\tilde{\bm\xi}'&,\omega_n)e^{i\tilde{\bm\xi'}\cdot\tilde{\bm\Pi}(\tilde{\bm R})/\alpha}\overline{\tilde{\Delta}}_{\sigma\sigma'}(\tilde{\bm R})\overline{\tilde{G}}^0_{\sigma'\sigma'}(\tilde{\bm\xi}',-\omega_n).
\end{align}
We then let $\tilde{\bm\xi}'=\alpha\tilde{\bm\xi}$, and obtain
\begin{align}
\overline{\tilde{\Delta}}_{\sigma\sigma'}(\tilde{\bm R})&=\frac{m^23V_0}{(2\pi)^2}\int\frac{d^3\tilde{\bm\xi}}{\tilde{\xi}^2}(\hat{\tilde{\xi}}_3\cos\theta'-\hat{\tilde{\xi}}_2\sin\theta')^2\nonumber\\
&\quad\times T\sum_{\omega_n}e^{-2|\omega_n|\tilde{\xi}/v_F}e^{i\tilde{\bm\xi}\cdot\tilde{\bm{\Pi}}(\tilde{\bm R})}\overline{\tilde{\Delta}}_{\sigma\sigma'}(\tilde{\bm R}),\label{finalsolution}
\end{align}
which is exactly as for an isotropic Fermi surface, except for the transformed $p$-wave polar/CBS state interaction and the modification of $\tilde{\bm{\Pi}}(\tilde{\bm R})$ due to $\alpha$ in Eq.~(\ref{Pi}). Note that in deriving Eq.~(\ref{finalsolution}), we used Eq.~(\ref{realspaceG0}) with $\tilde{\bm r}-\tilde{\bm r}'\rightarrow \alpha\tilde{\bm \xi}$. Since this form appears to describe the interaction in real space rather than in the correct momentum space, we rewrite this equation including the $\hat{\tilde{\bm k}}$ or $\hat{\tilde{\bm k}}'$ dependence of the order parameter, and also include the pairing interaction. $N(0)$, the single-spin density of states, can also be included in the expression by letting $\tilde{\bm \xi}\rightarrow\tilde{\bm k}'v_F$. We then obtain the expression in terms of the general transformed interaction $\tilde{V}(\hat{\tilde{\bm k}},\hat{\tilde{\bm k}}')$,
\begin{align}
\overline{\tilde{\Delta}}(\tilde{\bm R},\hat{\tilde{\bm k}})&=T\frac{N(0)}{2}\sum_{\omega_n}\int d\Omega_{\tilde{\bm k}'}\tilde{V}(\hat{\tilde{\bm k}},\hat{\tilde{\bm k}}')\int_0^{\infty}d\xi_{\tilde{\bm k}'}\notag\\
&\quad\times e^{-2\xi_{\tilde{\bm k}'}|\omega_n|}e^{-i\xi_{\tilde{\bm k}'}v_F\hat{\tilde{\bm k}'}\cdot\tilde{\bm\Pi}(\tilde{\bm R})}\overline{\tilde{\Delta}}(\tilde{\bm R},\hat{\tilde{\bm k}}'),
\end{align}
where $\tilde{V}(\hat{\tilde{\bm k}},\hat{\tilde{\bm k}})$ for the polar state with completely broken symmetry is given by Eq.~(\ref{tildeV}), but can be generalized to any anisotropic form. Of course, for non-parallel spin states, the Zeeman energies leading to Pauli pairbreaking and ${\bm B}$ at an arbitrary direction must also be included and properly transformed for an ellipsoidal FS.

We note that $\tilde{\bm B}=\hat{\bm z}\tilde{B}_3$. Neglecting defects and surface pinning effects, it is valid just below $B_{c2}$ to assume straight vortices along $\hat{\tilde{\bm z}}$. For a spatially constant (single-ferromagnetic domain) $\tilde{B}_3$, the $\tilde{\bm A}(\tilde{\bm R})$ can then be chosen to be either $-\tilde{B}_3\hat{\tilde{\bm X}}\tilde{Y}$ or $\tilde{B}_3\hat{\tilde{\bm Y}}\tilde{X}$, mapping the eigenvalue problem onto that of a one-dimensional (1D) harmonic oscillator.

In order to calculate $B_{c2}$, we expand $\overline{\tilde{\Delta}}(\tilde{\bm R},\hat{\tilde{\bm k}})$ in terms of the $\hat{\tilde{\bm k}}$ factor in $\tilde{V}(\hat{\tilde{\bm k}},\hat{\tilde{\bm k}}')$ and the $\tilde{\bm R}$ part in terms of the 1D harmonic oscillator eigenfunctions \cite{SK1980,HW},
\begin{align}
\overline{\tilde{\Delta}}(\tilde{\bm R},\hat{\tilde{\bm k}})=(\hat{\tilde{\bm k}}\cos\theta'-\hat{\tilde{\bm k}}\sin\theta')\sum_{n=0}^{\infty}a_n|n(\tilde{\bm R})\rangle.
\end{align}
The procedure is precisely the same as for the polar, SK and polar/CBS states \cite{SK1980,SK1985}, with the only differences being the $\theta'$ of the transformed interaction and the modification of the the operator from ${\bm \Pi}({\bm R})\rightarrow\tilde{\bm\Pi}(\tilde{\bm R})$, where $\tilde{\bm\Pi}(\tilde{\bm R})$ is given by Eq.~(\ref{Pi}) of the text. As in those previous calculations \cite{HW,SK1980,SK1985}, one requires the matrix elements
\begin{align}
M_{n',n}=\langle n'(\tilde{\bm R})|e^{i\xi_{\tilde{\bm k}'}v_F\hat{\tilde{\bm k}}'\cdot\tilde{\bm\Pi}(\tilde{\bm R})}|n(\tilde{\bm R})\rangle,
\end{align}
which must then be integrated over $\xi_{\tilde{\bm k}'}$ and the angles arising from $\hat{\tilde{\bm k}}'\cdot\tilde{\bm R}$. We write
\begin{align}
\tilde{\Pi}_{\pm}(\tilde{\bm R})=\frac{1}{\sqrt{2}}[\tilde{\Pi}_x(\tilde{\bm R})\pm i\tilde{\Pi}_y(\tilde{\bm R})],
\end{align}
and since $\tilde{\bm B}=\hat{\tilde{\bm z}}\tilde{B}_3$ is along the transformed $\tilde{z}$ axis, we may write
\begin{align}
&e^{i\xi_{\tilde{\bm k}'}v_F\hat{\tilde{\bm k}}'\cdot\tilde{\bm\Pi}(\tilde{\bm R})}\notag\\
=\,&e^{-\frac{1}{2}eB\alpha v_F^2\xi^2_{\tilde{\bm k}'}}e^{\frac{i}{\sqrt{2}}v_F\xi_{\tilde{\bm k}'}\sin\theta_{\tilde{\bm k}'}e^{-i\phi_{\tilde{\bm k}'}}\tilde{\Pi}_{+}(\tilde{\bm R})}\notag\\
&\times e^{\frac{i}{\sqrt{2}}v_F\xi_{\tilde{\bm k}'}\sin\theta_{\tilde{\bm k}'}e^{+i\phi_{\tilde{\bm k}'}}\tilde{\Pi}_{-}(\tilde{\bm R})}e^{iv_F\xi_{\tilde{\bm k}'}\cos\theta_{\tilde{\bm k}'}\tilde{\Pi}_z(\tilde{\bm R})}.
\end{align}
 For straight vortices, $\tilde{\Pi}_z(\tilde{\bm R})|n(\tilde{\bm R})\rangle=0$. Hence, we may drop the right factor containing $\tilde{\Pi}_z(\tilde{\bm R})$. Note that for this operator ordering, $\tilde{\Pi}^n_{-}(\tilde{\bm R})|n(\tilde{\bm R})\rangle=0$, etc. It is then easiest to expand the exponentials of the operators in the usual power series, and obtain the matrix elements
 \begin{align}
 M_{n',n}^{p',p}=\langle n'(\tilde{\bm R})|\tilde{\Pi}_{+}^{p'}(\tilde{\bm R})\tilde{\Pi}_{-}^p(\tilde{\bm R})|n(\tilde{\bm R})\rangle.
 \end{align} Then, one evaluates the integrals over $\theta_{\tilde{\bm k}'}$, $\phi_{\tilde{\bm k}'}$, and $\xi_{\tilde{\bm k}'}$ to obtain the relevant recursion relation for the $a_n$ coefficients.



\begin{thebibliography}{99}
\bibitem{Huy} N. T. Huy, A. Gasparini, D. E. de Nijs, Y. Huang, J. C. P. Klaasse, T. Gortenmulder, A. de Visser, A. Hamann, T. G{\"o}rlach, and H. v. L{\"o}hneysen, Phys. Rev. Lett. \textbf{99}, 067006 (2007).
\bibitem{deVisser} A. de Visser, N. T. Huy, A. Gasparini, D. E. de Nijs, D. Andreica, C. Baines, and A. Amato, Phys. Rev. Lett. {\bf 102}, 167003 (2009).
\bibitem{Aoki} D. Aoki, A. Huxley, E. Ressouche, D. Braithwaite, J. Flouquet, J.-P. Brison, E. Lhotel, and C. Paulsen, Nature \textbf{413},
613 (2001).
\bibitem{HH} F. Hardy and A. D. Huxley, Phys. Rev. Lett.\textbf{94}, 247006 (2005).
\bibitem{Levy} F. L\'{e}vy, I. Sheikin, and A. Huxley, Nature Phys. {\bf 3}, 460 (2007).
\bibitem{Yelland} E. A. Yelland, J. M. Barraclough, W. Wang, K. V. Kamenev, and A. D. Huxley, Nature Phys. {\bf 7}, 890 (2011).
\bibitem{Levy2} F. L{\'e}vy, I. Sheikin, B. Grenier, C. Marcenat, and A. Huxley, J. Phys.: Condens. Matter {\bf 21}, 164211 (2009).
\bibitem{Aoki2} D. Aoki, T. D. Matsuda, V. Taufour, E. Hassinger, G. Knebel, and J. Flouquet, J. Phys. Soc. Jpn. {\bf 80}, 013705 (2011).
\bibitem{AokiFlouquet} D. Aoki and J. Flouquet, J. Phys. Soc. Jpn. {\bf 81}, 011003 (2012).
\bibitem{SK1980} K. Scharnberg and R. A. Klemm, Phys. Rev. B \textbf{22}, 5233 (1980).
\bibitem{SK1985} K. Scharnberg and R. A. Klemm, Phys. Rev. Lett. \textbf{54}, 2445 (1985).
\bibitem{KS} R. A. Klemm and K. Scharnberg, Phys. Rev. B {\bf 24}, 6361 (1981).
\bibitem{Mineev} V. P. Mineev, C. R. Physique {\bf 7}, 35 (2006).
\bibitem{Davis} M. Divi{\v s}, L. M. Sandratskii, M. Richter, P. Mohn, and P. Nov{\'a}k, J. Alloys Comp. {\bf 337}, 48 (2002).
\bibitem{Shick} A. B. Shick, Phys. Rev. B. {\bf 65}, 180509(R) (2002).
\bibitem{Mueller} W. M{\"u}ller, V. H. Tran, and M. Richter, Phys. Rev. B {\bf 80}, 195108 (2009).
\bibitem{VG} G. E. Volovik and L. P. Gor\textquoteright{}kov,
Zh. Eksp. Teor. Fiz. \textbf{88}, 1412 (1985) [Sov. Phys.
JETP \textbf{61}, 843 (1985).]
\bibitem{Blount} E. I. Blount, Phys. Rev. B \textbf{32}, 2935 (1985).
\bibitem{Sauls} J. Sauls, Adv. Phys. \textbf{43}, 113 (1994).
\bibitem{Shivaram} B. S. Shivaram, Y. H. Jeong, T. F. Rosenbaum, and D. G. Hinks, Phys. Rev. Lett. {\bf 56}, 1078 (1986).
\bibitem{ChoiSauls} C. H. Choi and J. Sauls, Phys. Rev. Lett. {\bf 66}, 484 (1991).
\bibitem{MachidaMachida} Y. Machida, A. Itoh, K. Izawa, Y. Haga, E. Yamamoto, N. Kimura, Y. Onuki, Y. Tsutsumi, and K. Machida, Phys. Rev. Lett. {\bf 108}, 157002 (2012).
\bibitem{MS} V. P. Mineev and K. V. Samokhin, \textit{Introduction to Unconventional Superconductivity}(New York: Gordon and Breach 1999).
\bibitem{MM} A. P. Mackenzie and Y. Maeno, Rev. Mod. Phys. {\bf 75}, 6547 (2003).
\bibitem{Maeno} Y. Maeno, S. Kittaka, T. Nomura, S. Yonezawa, and K. Ishida, J. Phys. Soc. Jpn. {\bf 81}, 011009 (2012)
\bibitem{Deguchi} K. Deguchi, Z. Q. Mao, and Y. Maeno, J. Phys. Soc. Jpn. {\bf 73}, 1313 (2004).
\bibitem{Kittaka} S. Kittaka, T. Nakamura, Y. Aono, S. Yonezawa, K. Ishida, and Y. Maeno, Phys. Rev. B {\bf 80}, 174514 (2009).
\bibitem{Yonezawa} S. Yonezawa, T. Kajikawa, and Y. Maeno, Phys. Rev. Lett. {\bf 110}, 077003 (2013).
\bibitem{Suderow} H. Suderow, V. Crespo, I. Guillamon, S. Vieira, F. Servant, P. Lejay, J. P. Brison, and J. Flouquet, New. J. Phys. {\bf 11}, 093004 (2009).
\bibitem{Machida} K. Machida and M. Ichioka, Phys. Rev. B {\bf 77}, 184515 (2008).
\bibitem{Kriener} M. Kriener, K. Segawa, Z. Ren, S. Sasaki, and Y. Ando, {\it Phys. Rev. Lett.} {\bf 106}, 127004 (2011).
\bibitem{BayTI} T. V. Bay, T. Naka, Y. K. Huang, H. Luigjes, M. S. Golden, and A. de Visser, Phys. Rev. Lett. {\bf 108}, 057001 (2012).
\bibitem{BW} R. Balian and N. R. Werthamer, Phys. Rev. {\bf 131}, 1553 (1963).
\bibitem{Zhang} J. Zhang, C. L{\"o}rscher, Q. Gu, and R. A. Klemm, to be published.
\bibitem{AM} P. W. Anderson and P. Morel, Phys. Rev. {\bf 123}, 1911 (1961).
\bibitem{AB} P. W. Anderson and W. F. Brinkman, Phys. Rev. Lett. {\bf 30}, 1108 (1973).
\bibitem{LM} I. A. Luk'yanchuk and V. P. Mineev, Sov. Phys. JETP {\bf 66}, 1168 (1987).
\bibitem{Lorscher} C. L{\"o}rscher, J. Zhang, Q. Gu, and R. A. Klemm, to be published.
\bibitem{KC} R. A. Klemm and J. R. Clem, Phys. Rev. B {\bf 21}, 1868 (1980).
\bibitem{book} R. A. Klemm, \textit{Layered Superconductors Volume 1} (Oxford University Press, Oxford, UK and New York, NY 2012).
\bibitem{PC} M. Prohammer and J. P. Carbotte, Phys. Rev. B \textbf{42}, 2032 (1990).
\bibitem{HW} E. Helfand and N. R. Werthamer, Phys. Rev. {\bf 147}, 288 (1966).
\bibitem{Rieck} C. T. Rieck and K. Scharnberg, Physica B {\bf 163}, 670 (1990).
\bibitem{Hall} B. Hall (private communication).
\bibitem{Steglich} H. A. Vieyra, N. Oeschler, S. Seiro, H. S. Jeevan, C. Geibel, D. Parker, and F. Steglich, Phys. Rev. Lett. {\bf 106}, 207001 (2011).
\bibitem{AGD} A. A. Abrikosov, L. N. Gor'kov, and I. E. Dzaloshinskii, {\it Methods of Quantum Field Theory in Statistical Physics} (Dover Books on Physics, 1975).
\end{thebibliography}
\end{document}